\documentclass[journal,draftcls,onecolumn,12pt]{IEEEtran}

\usepackage{cite}
\usepackage{graphicx}
\usepackage[cmex10]{amsmath}
\usepackage{amsfonts}
\usepackage{amssymb}

\ifCLASSOPTIONcompsoc
\usepackage[caption=false,font=normalsize,labelfon
t=sf,textfont=sf]{subfig}
\else
\usepackage[caption=false,font=footnotesize]{subfig}
\fi

\usepackage{amsthm}

\usepackage{algorithm}
\usepackage{algorithmic}

\usepackage{tikz}
\usepackage{pgflibraryshapes}
\tikzstyle{block}=[draw opacity=0.7,line width=1.4cm]

\newtheorem{remark}{Remark}

\newtheorem{definition}{Definition}
\newtheorem{theorem}{Theorem}

\newtheorem{lemma}{Lemma}

\theoremstyle{definition}

\newcommand{\defeq}{\mbox {$  \ \stackrel{\Delta}{=} $}}

\newcommand{\be}{ \mbox{$ \bf E $}}

\theoremstyle{remark}

\begin{document}

\title{Non-asymptotic Equipartition Properties for Independent and Identically Distributed Sources
\thanks{This work was supported in part
by the Natural Sciences and Engineering Research Council of Canada
under Grant RGPIN203035-11, and by the Canada
Research Chairs Program.}}

\author{En-hui Yang and Jin Meng\thanks{En-hui Yang and Jin Meng are with the Dept.
of Electrical and Computer
Engineering, University of Waterloo, Waterloo, Ontario N2L 3G1,
Canada. Email: ehyang@uwaterloo.ca, j4meng@uwaterloo.ca} }

\date{October 20, 2011}
\maketitle

\newpage

\begin{abstract}
Given an independent and identically distributed source $X = \{X_i \}_{i=1}^{\infty}$ with finite Shannon entropy or differential entropy (as the case may be) $H(X)$, the
non-asymptotic equipartition property (NEP) with respect to $H(X)$ is established, which
characterizes, for any finite block length $n$, how close $-{1\over n}
\ln p(X_1 X_2 \cdots X_n) $ is to $H(X)$ by determining  the
information spectrum of $X_1 X_2 \cdots X_n $, i.e., the distribution
of $-{1\over n} \ln p(X_1 X_2 \cdots X_n) $.   Non-asymptotic
equipartition properties (with respect to conditional entropy, mutual
information, and relative entropy) in a similar nature are also
established. These non-asymptotic equipartition properties are
instrumental to the development of non-asymptotic coding (including
both source and channel coding) results in information theory in the
same way as the asymptotic equipartition property to all asymptotic
coding theorems established so far in information theory.  As an
example, the NEP with respect to $H(X)$ is used to establish a
non-asymptotic fixed rate source coding theorem, which reveals, for
any finite block length $n$, a complete picture about the tradeoff
between the minimum rate of fixed rate coding of $X_1 \cdots X_n$ and
error probability when the error probability is a constant, or goes to
$0$ with block length $n$ at a sub-polynomial $n^{-\alpha}$, $0 <
\alpha <1$, polynomial $n^{-\alpha}$, $ \alpha \geq 1$, or
sub-exponential $e^{- n^{\alpha}}$, $0< \alpha <1$, speed. In
particular, it is shown 
that for any finite block length $n$, the minimum rate (in nats per
symbol) of fixed rate coding of $X_1 X_2 \cdots X_n$ with error
probability $\Theta \left( \frac{n^{-\alpha}}{\sqrt{\ln n}} \right)$
is $H(X) + \sqrt{\sigma^2_H (X) (2
  \alpha )} \sqrt{\ln n \over n} +O( {\ln n \over n})$,
where $\alpha > 0$ and $\sigma^2_H (X) = {\bf E} [-\ln
p(X_1) ]^2 - H^2 (X) $ is the information variance of $X$.  With the help of the NEP with respect to
other information quantities, non-asymptotic channel coding theorems of similar nature will be established in a separate paper.

\end{abstract}


\newpage

\section{Introduction}
\label{sec0}
\setcounter{equation}{0}

Consider an independent and identically distributed (IID) source $X = \{X_i \}_{i=1}^{\infty}$ with source alphabet $\cal X$ and finite entropy $H(X) $, where $H(X)$ is the Shannon entropy  of $X_i$ if $\cal X$ is discrete, and the differential entropy of $X_i$ if $\cal X$ is the real line and each $X_i$ is a continuous random variable. Let $p(x)$ be the probability mass function (pmf) or probability density function (pdf) (as the case may be) of $X_i$.  The asymptotic equipartition property (AEP) for $X$ is the assertion that
 \begin{equation} \label{eq0-1}
  - {1 \over n} \ln p(X_1 X_2 \cdots X_n) \to H(X)
  \end{equation}
   either in probability or with probability one as $n$ goes to $\infty$. It implies that for sufficiently large $n$, with high probability, the outcomes of $X_1 X_2 \cdots X_n$ are approximately equiprobable with their respective probability ranging from $ e^{-n (H(X) + \epsilon)}$ to $e^{-n (H(X) - \epsilon)}$, where $\epsilon >0$ is a small fixed number. Here and throughout the rest of the paper, $\ln$ stands for the logarithm with base $e$, and all information quantities are measures in nats.

 The AEP is fundamental to information theory. It is not only instrumental to lossless source coding theorems, but also behind almost all asymptotic coding (including source, channel, and multi-user coding) theorems through the concepts of typical sets and typical sequences \cite{Cover}.

 However, in the non-asymptotic regime where one wants to establish non-asymptotic coding results for finite block length $n$, the AEP in its current form can not be applied in general. In this paper, we aim to establish the non-asymptotic counterpart of the AEP, which is broadly referred to as the non-asymptotic equipartition property (NEP), so that the NEP can be applied to finite block length $n$. Specifically, with respect to $H(X)$, we first characterize, for any finite block length $n$,
 how close $-{1\over n} \ln p(X_1 X_2 \cdots X_n) $ is to $H(X)$ by determining  the information spectrum of $X_1 X_2 \cdots X_n $, i.e., the distribution of $-{1\over n} \ln p(X_1 X_2 \cdots X_n) $; such a property is referred to as the NEP with respect to $H(X)$.   For any IID source pair $(X, Y) = \{(X_i, Y_i )\}_{i=1}^{\infty}$ with finite conditional entropy $H(X|Y) $ and mutual information $I(X; Y)$, where $H(X|Y)$ is the Shannon conditional entropy of $X_i$ given $Y_i$ if $X$ is discrete, and the  conditional differential entropy of $X_i$ given $Y_i$ if $X$ is continuous, we then examine, for any finite block length $n$,  how close $-{1\over n} \ln p(X^n |Y^n)$ ($- {1\over n} \ln {p(Y^n |X^n) \over p(Y^n) } $, respectively) is to $H(X|Y)$ ($I(X; Y)$, respectively) by determining the distribution of  $-{1\over n} \ln p(X^n |Y^n)$ ($- {1\over n} \ln {p(Y^n |X^n) \over p(Y^n) } $, respectively), where $p(x^n |y^n)$ ($p(y^n |x^n)$, respectively) is the conditional pmf or pdf (as the case may be) of $x^n =x_1 x_2 \cdots x_n $ ($y^n=y_1 y_2 \cdots y_n$, respectively) given $y^n$ ($x^n$, respectively); these properties are referred to as the NEP with respect to $H(X|Y)$ and $I (X;Y)$, respectively.

 In the same way as the AEP plays an important role in establishing the asymptotic
 coding (including source, channel, and multi-user coding) results in
 information theory, our established NEP is also instrumental to the
 development of non-asymptotic source and channel coding
 results. Using the NEP with respect to $H(X)$, we further establish a non-asymptotic fixed rate source coding theorem, which reveals, for any finite block length $n$, a complete picture about the tradeoff between the minimum rate of fixed rate coding of $X_1 \cdots X_n$ and error probability when the error probability is a constant, or goes to $0$ with block length $n$ at a sub-polynomial $n^{-\alpha}$, $0 < \alpha <1$, polynomial $n^{-\alpha}$, $ \alpha \geq 1$, or sub-exponential $e^{- n^{\alpha}}$, $0< \alpha <1$, speed. In particular, it is shown that
 for any finite block length $n$, the minimum rate (in nats per
 symbol) of fixed rate coding of $X_1 X_2 \cdots X_n$ with error
 probability $\Theta \left( \frac{n^{-\alpha}}{\sqrt{\ln n}} \right)$
 is  $H(X) + \sqrt{\sigma^2_H (X) (2
   \alpha )} \sqrt{\ln n \over n} +O( {\ln n \over n}) $, where
 $\alpha > 0$ and $\sigma^2_H (X) = {\bf E} [-\ln
 p(X_1) ]^2 - H^2 (X) $ is the information variance of $X$.  In a
 separate paper \cite{jar_decoding}, non-asymptotic channel coding theorems of similar nature will be established  with the help of the
 NEP with respect to other information quantities; in particular, it is shown \cite{jar_decoding} that for any binary
 input memoryless channel with uniform capacity achieving input $X$,
 random linear codes of block length $n$ can reach within
 $\sqrt{\sigma^2_H (X|Y) (2 \alpha )} \sqrt{\ln n \over n} +O( {\ln n
   \over n})$ of the channel capacity  while maintaining word error
 probability $\Theta \left( \frac{n^{-\alpha}}{\sqrt{\ln n}} \right)$,  where $\alpha >0$ and $  \sigma^2_H
 (X|Y) = {\bf E} [-\log p(X|Y) ]^2 - H^2 (X |Y) $ is the conditional
 information variance of $X$ given $Y$ with $Y$ being the output of
 the channel in response to the input $X$.

 The rest of the paper is organized as follows. Section \ref{sec1} is
 devoted to the NEP with respect to $H(X)$. All results in Section
 \ref{sec1} are then extended to the case of $H(X|Y)$ in Section
 \ref{sec2}, thereby establishing the NEP with respect to $H(X|Y)$. In
 Section \ref{sec3}, we analyze the NEP with respect to the mutual
 information and relative entropy. Finally, in Section \ref{sec4}, we
 apply the NEP with respect to $H(X)$ to investigate the performance
 of optimal fixed rate coding of $X_1 X_2 \cdots X_n$.

 \section{NEP With Respect to Entropy}
\label{sec1}
\setcounter{equation}{0}

 Define
\begin{equation} \label{eq1}
 \lambda^* (X) \defeq \sup \left\{\lambda \geq 0: \int p^{-\lambda +1} (x)  d x < \infty \right \}
\end{equation}
where $\int d x $ is understood throughout this paper to be the summation over the source alphabet of $X$  if $X$ is discrete.  Suppose that
\begin{equation} \label{eq1+}
 \lambda^* (X) >0\;.
\end{equation}
Let
\begin{equation} \label{eq2}
\sigma^2_H (X)  \defeq \int p(x) [-\ln p(x) ]^2 d x - H^2 (X)
\end{equation}
which will be referred to as the information variance of $X$. It is not hard to see that under the assumption \eqref{eq1+},
\begin{equation} \label{eq2+}
\int {p^{-\lambda +1} (x) \over \left [ \int p^{-\lambda+1} (y) d y \right ] } \left |- \ln p(x) \right |^k d x < \infty
\end{equation}
and
\[ \int p^{-\lambda +1} (x)  d x < \infty \]
 for any $\lambda \in (0, \lambda^* (X))$ and any positive integer $k$.
Further assume that
\begin{equation} \label{eq1+++}
  \sigma^2_H (X) >0 \mbox{ and } \int p(x) | \ln p(x) |^3 d x < \infty \;.
\end{equation}
 Then we have the following result, which will be referred to as the weak right NEP with respect to $H(X)$.

\begin{theorem}[Weak Right NEP] \label{th1}
For any $\delta  \geq 0$, let
 \[ r_{X} (\delta) \defeq \sup_{\lambda \geq 0} \left [ \lambda (H(X) + \delta) -
 \ln \int p^{-\lambda +1} (x) d x \right ] \;. \]
Then the following hold:
\begin{description}
\item[(a)] For any positive integer $n$,
  \begin{equation} \label{eq3}
  \Pr \left \{ - {1 \over n} \ln p(X^n ) > H(X) + \delta \right \} \leq e^{- n r_{X} (\delta) }
  \end{equation}
  where $X^n = X_1 X_2 \cdots X_n$.
 \item[(b)] Under the assumptions \eqref{eq1+} and \eqref{eq1+++},  there exists a $\delta^* >0$ such that for any $\delta \in (0, \delta^*]$ and any positive integer $n$,
 \begin{equation} \label{eq4-}
 r_{X} (\delta) = {1 \over 2 \sigma^2_H (X) } \delta^2 + O(\delta^3)
  \end{equation}
  and hence
 \begin{equation} \label{eq4}
  \Pr \left \{ - {1 \over n} \ln p( X^n ) > H(X) + \delta \right \} \leq e^{- n ({\delta^2 \over 2\sigma^2_H (X) } + O(\delta^3) )}\;.
  \end{equation}
  \end{description}

\end{theorem}

\begin{IEEEproof}[Proof of Theorem~\ref{th1}]
 The inequality \eqref{eq3} follows from the Chernoff bound. To see this is indeed the case, note that
 \begin{eqnarray}
 \Pr \left \{ - {1 \over n} \ln p(X_1 X_2 \cdots X_n ) > H(X) + \delta \right \}
 & = & \Pr \left \{ -  \ln p(X_1 X_2 \cdots X_n )  > n ( H(X) + \delta)  \right \}
 \nonumber \\
 & \leq & \inf_{ \lambda \geq 0} { {\bf E} [ e^{- \lambda \ln p(X_1 X_2 \cdots X_n )} ] \over e^{ n \lambda ( H(X) + \delta)} } \nonumber \\
 & = &  \inf_{ \lambda \geq 0} e^{-n \left [ \lambda ( H(X) + \delta) - \ln {\bf E} [ p^{-\lambda} (X_1) ] \right ] } \nonumber \\
 & = &  \inf_{ \lambda \geq 0} e^{-n \left [ \lambda ( H(X) + \delta) - \ln \int p^{-\lambda +1} (x) d x \right ] } \nonumber \\
 & = & e^{-n r_X (\delta) }\;. \label{eq5}
 \end{eqnarray}

 To show \eqref{eq4-} and \eqref{eq4}, we first analyze the property of $ r_X (\delta) $ as a function of $\delta$ over the region $\delta \geq 0$. It is easy to see that $ r_X (\delta) $ is convex and non-decreasing.  For any $\lambda \in [0, \lambda^* (X))$, define
 \begin{equation} \label{eq6}
 \delta(\lambda) \defeq \int {p^{-\lambda +1} (x) \over \left [ \int p^{-\lambda +1} (y) d y \right ] } \left [ - \ln p(x) \right ] d x - H(X)
 \end{equation}
 which, in view of \eqref{eq2+}, is well defined. Using a similar argument as in \cite[Properties 1 to 3]{yang-redundancy}, it is not hard to show that under the assumption \eqref{eq1+}, $\delta(\lambda)$ as a function of $\lambda$ is continuously differentiable up to any order over $\lambda \in (0, \lambda^* (X))$. Taking the first order derivative of $\delta (\lambda)$ yields
   \begin{eqnarray} \label{eq7}
     \delta' (\lambda) &=   &   \int {p^{-\lambda +1} (x) \over \left [ \int p^{-\lambda+1} (y) d y \right ] } \left [ - \ln p(x) \right ]^2 d x    - \left [ \int {p^{-\lambda +1} (x) \over \left [ \int p^{-\lambda +1} (y) d y \right ] } \left [ - \ln p(x) \right ] d x  \right ]^2     \nonumber \\
     & > & 0
       \end{eqnarray}
 where the last inequality is due to \eqref{eq1+++}. It is also easy to see that $\delta (0) =0$ and $\delta'(0) = \sigma^2_H (X)$. Therefore, $\delta (\lambda)$ is strictly increasing over $\lambda \in [0, \lambda^* (X))$. On the other hand, it is not hard to verify that under the assumption \eqref{eq1+}, the function $\lambda (H(X) + \delta) -\ln \int p^{-\lambda +1} (x) d x $ as a function of $\lambda$ is continuously differentiable over $\lambda \in [0, \lambda^* (X))$ with its derivative equal to
  \begin{equation} \label{eq7+}
   \delta - \delta(\lambda) \;.
  \end{equation}
  To continue, we distinguish between two cases: (1) $\lambda^* (X) = \infty$, and (2) $\lambda^* (X) < \infty$. In case (1), since $\delta (\lambda)$ is strictly increasing over $\lambda \in [0, \infty)$, it follows that for any $\delta = \delta (\lambda)$ for some $\lambda \in [0, \lambda^* (X))$, the supremum in the definition of $r_X (\delta)$ is actually achieved at that particular $\lambda$, i.e.,
 \begin{equation} \label{eq8}
 r_X (\delta(\lambda)) = \lambda (H(X) + \delta(\lambda) ) -
 \ln \int p^{-\lambda +1} (x) d x \;.
  \end{equation}
  In case (2), we have that for any $\delta = \delta (\lambda)$ for some $\lambda \in [0, \lambda^* (X))$ ,
   \begin{equation} \label{eq8+}
 \beta (H(X) + \delta(\lambda) ) -
 \ln \int p^{-\beta +1} (x) d x  < \lambda (H(X) + \delta(\lambda) ) -
 \ln \int p^{-\lambda +1} (x) d x
  \end{equation}
 for any $\beta \in [0, \lambda^* (X))$ with $\beta \not = \lambda$. In view of the definition of $\lambda^* (X)$, \eqref{eq8+} remains valid for any $\beta > \lambda^* (X)$ since then the left side of \eqref{eq8+} is $-\infty$. What remains to check is when $\beta = \lambda^* (X)$. If
\[ \int p^{-\lambda^* (X) +1} (x)  d x = \infty \]
it is easy to see that \eqref{eq8+} holds as well when $\beta = \lambda^* (X)$.  Suppose now
\[ \int p^{-\lambda^* (X) +1} (x)  d x < \infty  \;.\]
In this case, it follows from the dominated convergence theorem that
 \[ \lim_{\beta \uparrow \lambda^{*} (X) }  \int p^{-\beta +1}(x) d x =
     \int p^{-\lambda^* (X) +1}(x) d x \]
  and hence by letting $\beta$ go to $\lambda^* (X)$ from the left, we see that
  \eqref{eq8+} holds as well when $\beta = \lambda^* (X)$. Putting all cases together, we always have that
  for any $\delta = \delta (\lambda)$ for some $\lambda \in [0, \lambda^* (X))$,
   \begin{equation} \label{eq8++}
 r_X (\delta(\lambda)) = \lambda (H(X) + \delta(\lambda) ) -
 \ln \int p^{-\lambda +1} (x) d x \;.
  \end{equation}

  Let
  \[ \Delta^* (X) \defeq \lim_{\lambda \uparrow \lambda^* (X)} \delta (\lambda) \;.\]
 Since both $\delta (\lambda)$ and $ \ln \int p^{-\lambda +1} (x) d x $ are continuously differentiable with respect to $\lambda \in (0, \lambda^* (X))$ up to any order, it follows from \eqref{eq8++} that $r_X (\delta)$ is also continuously differentiable with respect to $\delta \in (0,
 \Delta^* (X))$ up to any order. (At $\delta =0$, $r_X (\delta)$ is continuously differentiable up to at least the third order inclusive.) Taking the first and second order derivatives of $r_X (\delta)$ with respect to $\delta$, we have
 \begin{eqnarray} \label{eq9}
r'_{X} (\delta) & = &  {d r_X (\delta) \over d \delta } \nonumber \\
 & = & {d r_X (\delta(\lambda)) \over d \lambda }  {d \lambda \over d \delta } \nonumber \\
 & = & {d r_X (\delta(\lambda)) \over d \lambda }  {1  \over \delta' (\lambda) } \nonumber \\
 & = & {1  \over \delta' (\lambda) }  \left [ H(X) + \delta (\lambda) + \lambda \delta' (\lambda)  -  \int {p^{-\lambda +1} (x) \over \left [ \int p^{-\lambda +1} (y) d y \right ] } \left [ - \ln p(x) \right ] d x  \right ] \nonumber \\
 & = & \lambda
 \end{eqnarray}
 and
 \begin{eqnarray} \label{eq10}
r''_{X} (\delta) & = & {d \lambda \over d \delta } \nonumber \\
    & = & { 1 \over \delta' (\lambda) }
  \end{eqnarray}
where $\delta = \delta (\lambda)$.  Therefore, $r_X (\delta)$ is convex, strictly increasing, and continuously differentiable up to at least the third order (inclusive) over $\delta \in [0, \Delta^* (X))$. Note that from \eqref{eq9} and \eqref{eq10}, we have $r'_X (0) = 0 $ and $r''_X (0) = 1/\sigma^2_H (X)$. Expanding $r_X (\delta)$ at $\delta =0$ by the Taylor expansion, we then have that there exists a $\delta^* >0$ such that
 \begin{equation} \label{eq11}
 r_X (\delta) = {1 \over 2 \sigma^2_H (X)} \delta^2 + O(\delta^3)
 \end{equation}
 for $\delta \in (0, \delta^*]$.  The inequality \eqref{eq4} now follows immediately from \eqref{eq3} and \eqref{eq11}. This completes the proof of Theorem~\ref{th1}.
\end{IEEEproof}

Having analyzed the function $r_X (\delta)$, we are now ready for a stronger version of the right NEP. 
For any $\lambda \in [0, \lambda^* (X))$, define
  \begin{equation} \label{eq12}
  f_{\lambda} (x) \defeq {p^{-\lambda} (x) \over \int p^{-\lambda +1} (y) d y }
  \end{equation}
   \begin{equation} \label{eq13}
  \sigma^2_H (X, \lambda)  \defeq \int  f_{\lambda} (x) p(x) \left | -\ln p(x) - (H(X) + \delta (\lambda) ) \right |^2 d x
  \end{equation}
\begin{equation} \label{eq14}
  M_H (X, \lambda)  \defeq \int  f_{\lambda} (x) p(x) \left | -\ln p(x) - (H(X) + \delta (\lambda) ) \right |^3 d x
  \end{equation}
and
\begin{equation} \label{eq15}
 f_{\lambda} (x^n) \defeq \prod_{i=1}^n f_{\lambda} (x_i)
  \end{equation}
where $\delta (\lambda)$ is defined in \eqref{eq6}. Write $M_H (X, 0)$ as $M_H(X)$. It is easy to see that $\sigma^2_H (X, 0) = \sigma^2_H (X)$, $ \sigma^2_H (X, \lambda) = \delta' (\lambda) $, and
\begin{equation} \label{eq15+}
M_H (X) = \int   p(x) \left | -\ln p(x) - H(X)  ) \right |^3 d x \; .
\end{equation}
 Then we have the following stronger result.

\begin{theorem}[Strong Right NEP] \label{th1+}
Under the assumptions \eqref{eq1+} and \eqref{eq1+++},  the following hold:
 \begin{description}
 \item[(a)] For any $\delta \in (0, \Delta^* (X))$ and any positive integer $n$
 \begin{equation} \label{eq16}
  \bar{\xi}_H(X,\lambda,n) e^{- n r_X (\delta) } \geq
  \Pr \left \{ - {1 \over n} \ln p(X^n ) > H(X) + \delta \right \} \geq
  \underline{\xi}_H (X,\lambda,n) e^{- n r_X (\delta) }
  \end{equation}
where $\lambda = r'_X (\delta) >0$, 
\begin{eqnarray}
  \label{eq16+}
  \lefteqn{\bar{\xi}_H(X,\lambda,n) = \frac{2CM_H(X,\lambda)}{\sqrt{n} \sigma^3_H(X,\lambda)}} \nonumber \\
                      &&{+}\: e^{\frac{n \lambda^2 \sigma^2_H(X,\lambda)}{2}} 
                      \left[ Q(\sqrt{n}  \lambda \sigma_H(X,\lambda)) - Q(\rho^*+\sqrt{n}  \lambda \sigma_H(X,\lambda))\right]
                      \\
  \label{eq16++}
  \lefteqn{\underline{\xi}_H(X,\lambda,n) = 
       e^{\frac{n \lambda^2 \sigma^2_H(X,\lambda)}{2}} Q(\rho_*+\sqrt{n}  \lambda \sigma_H(X,\lambda))}
\end{eqnarray}
with $Q(\rho^*) = \frac{CM_H(X,\lambda)}{\sqrt{n} \sigma^3_H(X,\lambda)}$ and 
$Q(\rho_*) = \frac{1}{2} - \frac{2CM_H(X,\lambda)}{\sqrt{n} \sigma^3_H(X,\lambda)}$,  
$Q(t) = {1\over \sqrt{2\pi}} \int_{t}^{\infty} e^{-u^2/2} du $ and
$C <1$ is the universal constant in the central limit theorem of Berry and Esseen.
 \item[(b)] For any  $ \delta \leq c \sqrt{\ln n \over n} $, where $c < \sigma_H (X)$ is a constant,
  \begin{eqnarray} \label{eq17+}
    Q  \left ( {\delta \sqrt{n} \over \sigma_H (X)} \right ) - {C M_H (X) \over \sqrt{n} \sigma^3_H (X)}
    & \leq  &  \Pr \left \{ - {1 \over n} \ln p(X^n ) > H(X) + \delta \right \} \nonumber \\
  &  \leq &  Q  \left ( {\delta \sqrt{n} \over \sigma_H (X)} \right ) + {C M_H (X) \over \sqrt{n} \sigma^3_H (X)} .
     \end{eqnarray}
  \end{description}
\end{theorem}

\begin{IEEEproof}[Proof of Theorem~\ref{th1+}]
From \eqref{eq8++}, it follows that with $\lambda = r'_X (\delta) $
  \begin{equation} \label{eq18}
    r_X (\delta ) = \lambda (H(X) + \delta ) - \ln \int p^{-\lambda +1} (x) d x \;.
  \end{equation}
Then it is not hard to verify that
  \begin{eqnarray} \label{eq19}
    \lefteqn{ \Pr \left \{ - {1 \over n} \ln p(X^n ) > H(X) + \delta \right \}}
    \nonumber \\
    & = & \int \limits _{ - {1 \over n} \ln p(x^n ) > H(X) + \delta } p(x^n) d x^n \nonumber \\
    & = & \int \limits _{ - {1 \over n} \ln p(x^n ) > H(X) + \delta } f^{-1}_{\lambda} (x^n) f_{\lambda} (x^n) p(x^n) d x^n \nonumber \\
    & = & \int \limits _{ - {1 \over n} \ln p(x^n ) > H(X) + \delta } 
              e^{ -n  \left [ -{1 \over n} \lambda \ln p(x^n) - \ln \int p^{-\lambda +1} (y) d y \right ] }
              f_{\lambda} (x^n) p(x^n) d x^n \nonumber \\
    & = & \int \limits _{ - {1 \over n} \ln p(x^n ) > H(X) + \delta} 
              e^{-n  \left [ -{1 \over n} \lambda \ln p(x^n) - \lambda (H(X)+\delta) + r_X(\delta)  \right ] }  
              f_{\lambda} (x^n) p(x^n) d x^n   \nonumber \\
    & = & e^{- n r_X(\delta)} \int \limits _{ - {1 \over n} \ln p(x^n ) > H(X) + \delta} 
              e^{-n  \lambda \left [ -{1 \over n} \ln p(x^n) - (H(X)+\delta)  \right ] }
              f_{\lambda} (x^n) p(x^n) d x^n \nonumber \\
    & = & e^{-n r_X (\delta)} \int \limits _{ - {1 \over n} \ln p(x^n ) > H(X) + \delta} 
              e^{- \sqrt{n}  \lambda \sigma_H(X,\lambda) \frac{- \ln p(x^n) - n(H(X)+\delta)}{\sqrt{n} \sigma_H(X,\lambda)} }
              f_{\lambda} (x^n) p(x^n) d x^n \nonumber \\
    & = & e^{-n r_X (\delta)} \int\limits_{\rho > 0} \int\limits_{\frac{- \ln p(x^n) - n(H(X)+\delta)}{\sqrt{n} \sigma_H(X,\lambda)} = \rho} 
             e^{- \sqrt{n}  \lambda \sigma_H(X,\lambda) \rho} f_{\lambda} (x^n) p(x^n) d x^n d \rho \nonumber \\
    & = & e^{-n r_X (\delta)} \int\limits^{+\infty}_{0} e^{- \sqrt{n}  \lambda \sigma_H(X,\lambda) \rho} d (1-\bar{F}_n(\rho)) \nonumber \\
    & = & e^{-n r_X (\delta)} \left[ \bar{F}_n(0) - \int\limits^{+\infty}_{0} 
             \sqrt{n}  \lambda \sigma_H(X,\lambda)  e^{- \sqrt{n}  \lambda \sigma_H(X,\lambda) \rho} 
             \bar{F}_n(\rho) d \rho \right]
  \end{eqnarray}
where the last equality is due to integration by parts, 
  \begin{eqnarray*}
    \bar{F}_n(\rho) &\defeq& \Pr \left\{ \frac{- \ln p(Z^n) - n(H(X)+\delta)}{\sqrt{n} \sigma_H(X,\lambda)} > \rho \right\} \\
                  &=& \Pr \left\{ \sum^n_{i=1} \frac{- \ln p(Z_i) - (H(X)+\delta)}{\sqrt{n} \sigma_H(X,\lambda)} > \rho \right\}
  \end{eqnarray*}
and $\{Z_i\}^n_{i=1}$ are IID random variables with pmf or pdf (as the case may be) $f_{\lambda}(x)p(x)$. Let
\begin{eqnarray}
  \label{eq19-1}
  \xi_n & \defeq & \bar{F}_n(0) - \int\limits^{+\infty}_{0} 
             \sqrt{n}  \lambda \sigma_H(X,\lambda)  e^{- \sqrt{n}  \lambda \sigma_H(X,\lambda) \rho} 
             \bar{F}_n(\rho) d \rho \\
  \label{eq19-2}
            & = & \int\limits^{+\infty}_{0} 
             \sqrt{n}  \lambda \sigma_H(X,\lambda)  e^{- \sqrt{n}  \lambda \sigma_H(X,\lambda) \rho} 
             [\bar{F}_n(0) - \bar{F}_n(\rho)] d \rho 
\end{eqnarray}
At this point, we invoke the following central limit theorem of Berry and Esseen\cite[Theorem 1.2]{hall}.
\begin{lemma} \label{le1}
 Let $V_1, V_2, \cdots$ be  independent real random variables with zero means
and finite third moments, and set
 \[ \sigma_n^2 = \sum_{i=1}^n \be V_i^2 . \]
Then there exists a universal constant $C < 1$ such that for any $n \geq 1$,
 \[ \sup_{-\infty < t < +\infty} \left| \Pr \left\{ \sum_{i=1}^n V_i > \sigma_n t \right\}
  - Q (t) \right| \leq C \sigma_n^{-3} \sum_{i=1}^n \be |V_i |^3 . \]
\end{lemma}
Towards evaluating $\xi_n$, we can bound $\bar{F}_n(\rho)$ in terms of $Q(\rho)$,
by applying Lemma \ref{le1} to $\{ - \ln p(Z_i) - (H(X) +\delta) \}_{i=1}^n$.
Then for $\rho > 0$, we have
\begin{eqnarray}
  \label{eq19-3}
  \bar{F}_n(0) & \leq & Q(0) + \frac{CM_H(X,\lambda)}{\sqrt{n} \sigma^3_H(X,\lambda)} \nonumber \\
                & = & \frac{1}{2} + \frac{CM_H(X,\lambda)}{\sqrt{n} \sigma^3_H(X,\lambda)} \\
  \label{eq19-4}
  \bar{F}_n(\rho) & \geq & \left[ Q(\rho)-\frac{CM_H(X,\lambda)}{\sqrt{n} \sigma^3_H(X,\lambda)}\right]^+
\end{eqnarray}
and
\begin{eqnarray}
  \label{eq19-5}
  \bar{F}_n(0) - \bar{F}_n(\rho) 
    & \geq & \left[ Q(0)-\frac{CM_H(X,\lambda)}{\sqrt{n} \sigma^3_H(X,\lambda)} 
                           - \left(Q(\rho)+\frac{CM_H(X,\lambda)}{\sqrt{n} \sigma^3_H(X,\lambda)}\right) 
                  \right]^+ \nonumber\\
    & = & \left[ \frac{1}{2} - Q(\rho) - \frac{2 CM_H(X,\lambda)}{\sqrt{n} \sigma^3_H(X,\lambda)} \right]^+
\end{eqnarray}
where $[x]^+ = \max \{ x, 0\}$. Now plugging \eqref{eq19-3} and \eqref{eq19-4} into \eqref{eq19-1} yields
\begin{eqnarray}
  \label{eq20-1}
  \xi_n &\leq& \frac{1}{2} + \frac{CM_H(X,\lambda)}{\sqrt{n} \sigma^3_H(X,\lambda)} -
                       \int\limits^{+\infty}_{0} 
                       \sqrt{n}  \lambda \sigma_H(X,\lambda)  e^{- \sqrt{n}  \lambda \sigma_H(X,\lambda) \rho} 
                       \left[ Q(\rho)-\frac{CM_H(X,\lambda)}{\sqrt{n} \sigma^3_H(X,\lambda)}\right]^+ d \rho \nonumber \\
            &=&    \frac{1}{2} + \frac{CM_H(X,\lambda)}{\sqrt{n} \sigma^3_H(X,\lambda)} -
                       \int\limits^{\rho^*}_{0} 
                       \sqrt{n}  \lambda \sigma_H(X,\lambda)  e^{- \sqrt{n}  \lambda \sigma_H(X,\lambda) \rho} 
                       \left[ Q(\rho)-\frac{CM_H(X,\lambda)}{\sqrt{n} \sigma^3_H(X,\lambda)}\right] d\rho \nonumber \\
            &=&    \frac{1}{2} + \frac{CM_H(X,\lambda)}{\sqrt{n} \sigma^3_H(X,\lambda)} -
                       \int\limits^{\rho^*}_{0} 
                       \left[ Q(\rho)-\frac{CM_H(X,\lambda)}{\sqrt{n} \sigma^3_H(X,\lambda)}\right] 
                       d \left( - e^{- \sqrt{n}  \lambda \sigma_H(X,\lambda) \rho} \right) \nonumber \\
            &=&    \frac{2CM_H(X,\lambda)}{\sqrt{n} \sigma^3_H(X,\lambda)} +
                       \int\limits^{\rho^*}_{0} \frac{1}{\sqrt{2 \pi}} e^{-\frac{\rho^2}{2}}
                       e^{- \sqrt{n}  \lambda \sigma_H(X,\lambda) \rho}  
                       d\rho \nonumber \\
            &=&    \frac{2CM_H(X,\lambda)}{\sqrt{n} \sigma^3_H(X,\lambda)} +
                       \int\limits^{\rho^*}_{0} \frac{1}{\sqrt{2 \pi}} 
                       e^{-\frac{(\rho+\sqrt{n}  \lambda \sigma_H(X,\lambda))^2}{2}+\frac{n \lambda^2 \sigma^2_H(X,\lambda)}{2}}
                       d\rho \nonumber \\
            &=&    \frac{2CM_H(X,\lambda)}{\sqrt{n} \sigma^3_H(X,\lambda)} 
                       + e^{\frac{n \lambda^2 \sigma^2_H(X,\lambda)}{2}} 
                      \left[ Q(\sqrt{n}  \lambda \sigma_H(X,\lambda)) - Q(\rho^*+\sqrt{n}  \lambda \sigma_H(X,\lambda))\right]
                      \nonumber \\
            &=&    \bar{\xi}_H (X,\lambda,n)
\end{eqnarray}
where $Q(\rho^*) = \frac{CM_H(X,\lambda)}{\sqrt{n} \sigma^3_H(X,\lambda)}$, and meanwhile plugging \eqref{eq19-5}
into \eqref{eq19-2} yields
\begin{eqnarray}
  \label{eq20-2}
  \xi_n &\geq& \int\limits^{+\infty}_{0} 
                        \sqrt{n}  \lambda \sigma_H(X,\lambda)  e^{- \sqrt{n}  \lambda \sigma_H(X,\lambda) \rho} 
                        \left[ \frac{1}{2} - Q(\rho) - \frac{2 CM_H(X,\lambda)}{\sqrt{n} \sigma^3_H(X,\lambda)} \right]^+ d\rho \nonumber \\
            &=&     \int\limits^{+\infty}_{\rho_*} 
                        \sqrt{n}  \lambda \sigma_H(X,\lambda)  e^{- \sqrt{n}  \lambda \sigma_H(X,\lambda) \rho} 
                        \left[ \frac{1}{2} - Q(\rho) - \frac{2 CM_H(X,\lambda)}{\sqrt{n} \sigma^3_H(X,\lambda)} \right] d\rho \nonumber \\
            &=&     \int\limits^{+\infty}_{\rho_*} 
                        \left[ \frac{1}{2} - Q(\rho) - \frac{2 CM_H(X,\lambda)}{\sqrt{n} \sigma^3_H(X,\lambda)} \right] 
                        d\left( -e^{- \sqrt{n}  \lambda \sigma_H(X,\lambda) \rho} \right) \nonumber \\
            &=&     \int\limits^{+\infty}_{\rho_*} \frac{1}{\sqrt{2 \pi}} e^{-\frac{\rho^2}{2}}  e^{- \sqrt{n}  \lambda \sigma_H(X,\lambda) \rho} d\rho \nonumber \\
            &=&     e^{\frac{n \lambda^2 \sigma^2_H(X,\lambda)}{2}} Q(\rho_*+\sqrt{n}  \lambda \sigma_H(X,\lambda)) \nonumber \\
            &=&     \underline{\xi}_H (X,\lambda,n)
\end{eqnarray}
where $Q(\rho_*) = \frac{1}{2} - \frac{2CM_H(X,\lambda)}{\sqrt{n} \sigma^3_H(X,\lambda)}$. 
Combining \eqref{eq19} with \eqref{eq20-1} and \eqref{eq20-2} completes the proof of part (a) of Theorem \ref{th1+}.

Applying Lemma \ref{le1} to the IID sequence $\{-\ln p(X_i) - H(X)\}^n_{i=1}$, we get \eqref{eq17+}. This completes the proof
of Theorem \ref{th1+}.
\end{IEEEproof}

\begin{remark}
  \label{remark1}
  Note that $\lambda = r'_X(\delta) = \Theta (\delta)$.
  When $\lambda=\Omega(1)$ with respect to $n$,
  it can be easily verified that $\bar{\xi}_H (X,\lambda,n)$ and $\underline{\xi}_H (X,\lambda,n)$ are both on the order
  of $\frac{1}{\sqrt{n}}$, by applying well-known inequality
  \begin{displaymath}
    \frac{1}{t+t^{-1}} \frac{1}{\sqrt{2 \pi}} e^{-\frac{t^2}{2}} \leq Q(t) \leq \frac{1}{t} \frac{1}{\sqrt{2 \pi}} e^{-\frac{t^2}{2}}.
  \end{displaymath}
  Meanwhile, on one hand, it is easy to see that
  \begin{eqnarray*}
    \bar{\xi}_H(X,\lambda,n)  &\leq& e^{\frac{n \lambda^2 \sigma^2_H(X,\lambda)}{2}} Q(\sqrt{n}  \lambda \sigma_H(X,\lambda))
                      + \frac{2CM_H(X,\lambda)}{\sqrt{n} \sigma^3_H(X,\lambda)}.
  \end{eqnarray*}
  On the other hand,
  \begin{eqnarray*}
    \underline{\xi}_H(X,\lambda,n) 
                             &=& e^{\frac{n \lambda^2 \sigma^2_H(X,\lambda)}{2}} Q(\sqrt{n}  \lambda \sigma_H(X,\lambda)) -
                                     e^{\frac{n \lambda^2 \sigma^2_H(X,\lambda)}{2}}
                                     \int\limits^{\rho_*+\sqrt{n}\lambda\sigma_H(X,\lambda)}_{\sqrt{n}\lambda\sigma_H(X,\lambda)} 
                                     \frac{1}{\sqrt{2 \pi}} e^{-\frac{\rho^2}{2}} d \rho\\
                             &=& e^{\frac{n \lambda^2 \sigma^2_H(X,\lambda)}{2}} Q(\sqrt{n}  \lambda \sigma_H(X,\lambda)) - 
                                     e^{\frac{n \lambda^2 \sigma^2_H(X,\lambda)}{2}}
                                     \int\limits^{\rho_*}_{0}
                                     \frac{1}{\sqrt{2 \pi}} e^{-\frac{(\rho+\sqrt{n}\lambda\sigma_H(X,\lambda))^2}{2}} d \rho \\
                             &=&  e^{\frac{n \lambda^2 \sigma^2_H(X,\lambda)}{2}} Q(\sqrt{n}  \lambda \sigma_H(X,\lambda)) -
                                      \int\limits^{\rho_*}_{0} 
                                      \frac{1}{\sqrt{2 \pi}} e^{-\frac{\rho^2+2 \rho \sqrt{n}\lambda\sigma_H(X,\lambda)}{2}} d \rho \\
                             &\geq&  e^{\frac{n \lambda^2 \sigma^2_H(X,\lambda)}{2}} Q(\sqrt{n}  \lambda \sigma_H(X,\lambda)) -
                                      \int\limits^{\rho_*}_{0} 
                                      \frac{1}{\sqrt{2 \pi}} e^{-\frac{\rho^2}{2}} d \rho \\
                             &=&  e^{\frac{n \lambda^2 \sigma^2_H(X,\lambda)}{2}} Q(\sqrt{n}  \lambda \sigma_H(X,\lambda)) -
                                      \frac{2CM_H(X,\lambda)}{\sqrt{n} \sigma^3_H(X,\lambda)}.
  \end{eqnarray*}
  To further shed light on $\bar{\xi}_H(X,\lambda,n) $ and $\underline{\xi}_H(X,\lambda,n) $, 
  we observe that
  \begin{displaymath}
     \frac{1}{\sqrt{2 \pi} \sqrt{n} \lambda \sigma_H(X,\lambda)+\frac{1}{\sqrt{2 \pi} \sqrt{n} \lambda \sigma_H(X,\lambda)}}
     \leq e^{\frac{n \lambda^2 \sigma^2_H(X,\lambda)}{2}} Q(\sqrt{n}  \lambda \sigma_H(X,\lambda)) 
     \leq \frac{1}{\sqrt{2 \pi} \sqrt{n} \lambda \sigma_H(X,\lambda)} .
  \end{displaymath}
  And therefore, whenever $\lambda = o(1)$ and $\lambda = \omega(n^{-1})$,
  \begin{displaymath}
     e^{\frac{n \lambda^2 \sigma^2_H(X,\lambda)}{2}} Q(\sqrt{n}  \lambda \sigma_H(X,\lambda)) 
      = \Theta \left( \frac{1}{\sqrt{n} \lambda} \right) = \omega \left( \frac{1}{\sqrt{n}} \right) 
  \end{displaymath}
  which further implies
  \begin{eqnarray*}
    \bar{\xi}_H(X,\lambda,n) &=& e^{\frac{n \lambda^2 \sigma^2_H(X,\lambda)}{2}} Q(\sqrt{n}  \lambda \sigma_H(X,\lambda))
    \left( 1 + o(1) \right) \\
    \underline{\xi}_H(X,\lambda,n) &=& e^{\frac{n \lambda^2 \sigma^2_H(X,\lambda)}{2}} Q(\sqrt{n}  \lambda \sigma_H(X,\lambda))
    \left( 1 - o(1) \right).
  \end{eqnarray*}
\end{remark}

\begin{remark}
  \label{remark2}
  Another interesting observation from the proof of Theorem \ref{th1+}, especially \eqref{eq19}, is the recursive relation
  between
  \begin{eqnarray*}
    \Pr \left \{ - {1 \over n} \ln p(X^n ) > H(X) + \delta \right \} 
         &=& \Pr \left \{ \frac{- \ln p(X^n ) - nH(X)}{\sqrt{n} \sigma_H(X)} > \frac{\delta}{\sqrt{n} \sigma_H(X)} \right \} \\
         &\defeq& \bar{F}_{X,n} \left( \frac{\delta}{\sqrt{n} \sigma_H(X)} \right)
  \end{eqnarray*}
  and
  \begin{displaymath}
   \bar{F}_{Z,n} (\rho) \defeq \bar{F}_n(\rho) = \Pr \left\{ \frac{- \ln p(Z^n) - n(H(X)+\delta)}{\sqrt{n} \sigma_H(X,\lambda)} > \rho \right\}.
  \end{displaymath}
  As shown in the proof, a proper bound on $\bar{F}_{Z,n} (\rho)$ (using Berry-Esseen Central Limit Theorem) results in
  a bound \eqref{eq16} on $\bar{F}_{X,n} \left( \frac{\delta}{\sqrt{n} \sigma_H(X)} \right)$. 
  To continue, we can apply this bound \eqref{eq16} on $\bar{F}_{Z,n} (\rho)$ to get another bound
  on $\bar{F}_{X,n} \left( \frac{\delta}{\sqrt{n} \sigma_H(X)} \right)$. 
  Numerically, we can keep tightening the bound on $\bar{F}_{X,n} \left( \frac{\delta}{\sqrt{n} \sigma_H(X)} \right)$ 
  in this recursive manner until no significant improvement can be made.
\end{remark}
  
  The probability that $-{1\over n} \ln p(X^n)$ is away from $H(X)$ to the left can be bounded similarly. Define
\begin{equation} \label{eql-1}
 \lambda^*_{-} (X) \defeq \sup \left\{\lambda \geq 0: \int p^{\lambda +1} (x)  d x < \infty \right \}\;.
\end{equation}
 Suppose that
\begin{equation} \label{eql-1+}
 \lambda^*_{-} (X) >0\;.
\end{equation}
  Define for any $\delta \geq 0$
  \[ r_{X, -} (\delta) \defeq \sup_{\lambda \geq 0} \left [ \lambda (\delta- H(X) ) -
 \ln \int p^{\lambda +1} (x) d x \right ]  \]
  and for any $\lambda \in [0, \lambda^*_{-} (X))$
   \[
 \delta_{-} (\lambda) \defeq \int {p^{\lambda +1} (x) \over \left [ \int p^{\lambda +1} (y) d y \right ] } \left [  \ln p(x) \right ] d x  + H(X) \;.
 \]
 Then under the assumption \eqref{eq1+++}, $\delta_{-} (\lambda)$ is strictly increasing over $    \lambda \in [0, \lambda^*_{-} (X))$ with   $\delta_{-} (0) =0$.
 Let
  \[ \Delta^*_{-} (X) = \lim_{\lambda \uparrow \lambda^*_{-} (X) } \delta_{-} (\lambda) \;. \]
 Following the proof of Theorem~\ref{th1}, we have that $r_{X, -} (\delta)$ is strictly increasing, convex, and continuously differentiable up to at least the third order inclusive over $\delta \in [0, \Delta^*_{-} (X))$, and furthermore
 \[  r_{X, -} (\delta) = \lambda (\delta- H(X) ) -
 \ln \int p^{\lambda +1} (x) d x  \]
 with $\lambda = r'_{X, -} (\delta) $ satisfying
    \[ \delta_{-} (\lambda) = \delta \;.\]
 Define
 \[
  \sigma^2_{H, -} (X, \lambda)  \defeq \int  {p^{\lambda +1} (x) \over \left [ \int p^{\lambda +1} (y) d y \right ] }   \left | -\ln p(x) - (H(X) - \delta_{-} (\lambda) ) \right |^2 d x
 \]
and
\[
  M_{H,-} (X, \lambda)  \defeq \int  {p^{\lambda +1} (x) \over \left [ \int p^{\lambda +1} (y) d y \right ] }   \left | -\ln p(x) - (H(X) - \delta_{-} (\lambda) ) \right |^3 d x \;.
\]
 In parallel with Theorems~\ref{th1} and \ref{th1+},  we have the following result, which is referred to as the left NEP with respect to $H(X)$ and can be proved similarly.
 \begin{theorem}[Left NEP] \label{th1++}
For any positive integer $n$,
  \begin{equation} \label{eql-3}
  \Pr \left \{ - {1 \over n} \ln p(X^n ) \leq H(X) - \delta \right \} \leq e^{- n r_{X, -} (\delta) }\;.
  \end{equation}
 Furthermore,  under the assumptions \eqref{eql-1+} and \eqref{eq1+++},  the following also hold:
 \begin{description}

 \item[(a)] There exists a $\delta^* >0$ such that for any $\delta \in (0, \delta^*]$ and any positive integer $n$,
 \begin{equation} \label{eql-4-}
 r_{X,-} (\delta) = {1 \over 2 \sigma^2_H (X) } \delta^2 + O(\delta^3)
  \end{equation}
  and hence
 \begin{equation} \label{eql-4}
  \Pr \left \{ - {1 \over n} \ln p( X^n ) \leq H(X) - \delta \right \} \leq e^{- n ({\delta^2 \over 2\sigma^2_H (X) } + O(\delta^3) )}\;.
  \end{equation}

  \item[(b)] For any $\delta \in (0, \Delta^*_{-} (X))$ and any positive integer $n$
 \begin{equation} \label{eql-16}
  \bar{\xi}_{H,-} (X,\lambda,n) e^{- n r_{X,-} (\delta)} \geq 
  \Pr \left \{ - {1 \over n} \ln p(X^n ) \leq H(X) - \delta \right \}
   \geq \underline{\xi}_{H,-} (X,\lambda,n) e^{- n r_{X,-} (\delta)}
  \end{equation}
  where $\lambda = r'_{X,-} (\delta) >0$, and 
  \begin{eqnarray}
  \label{eql16+}
  \lefteqn{\bar{\xi}_{H,-}(X,\lambda,n) = \frac{2CM_{H,-}(X,\lambda)}{\sqrt{n} \sigma^3_{H,-}(X,\lambda)} } \nonumber \\
                      &&{+}\: e^{\frac{n \lambda^2 \sigma^2_{H,-}(X,\lambda)}{2}} 
                      \left[ Q(\sqrt{n}  \lambda \sigma_{H,-}(X,\lambda)) - Q(\rho^*+\sqrt{n}  \lambda \sigma_{H,-}(X,\lambda))\right] \\      
  \label{eql16++}
  \lefteqn{\underline{\xi}_{H,-}(X,\lambda,n) = 
           e^{\frac{n \lambda^2 \sigma^2_{H,-}(X,\lambda)}{2}} Q(\rho_*+\sqrt{n}  \lambda \sigma_{H,-}(X,\lambda))}
  \end{eqnarray}
  with $Q(\rho^*) = \frac{CM_{H,-}(X,\lambda)}{\sqrt{n} \sigma^3_{H,-}(X,\lambda)}$ and 
  $Q(\rho_*) = \frac{1}{2} - \frac{2CM_{H,-}(X,\lambda)}{\sqrt{n} \sigma^3_{H,-}(X,\lambda)}$.

 \item[(c)] For any  $ \delta \leq c \sqrt{\ln n \over n} $, where $c < \sigma_H (X)$ is a constant,
  \begin{eqnarray} \label{eql-17+}
    Q  \left ( {\delta \sqrt{n} \over \sigma_H (X)} \right ) - {C M_H (X) \over \sqrt{n} \sigma^3_H (X)}
    & \leq  &  \Pr \left \{ - {1 \over n} \ln p(X^n ) \leq H(X) - \delta \right \} \nonumber \\
  &  \leq &  Q  \left ( {\delta \sqrt{n} \over \sigma_H (X)} \right ) + {C M_H (X) \over \sqrt{n} \sigma^3_H (X)} \;.
     \end{eqnarray}
  \end{description}

\end{theorem}

Remarks similar to those (Remark \ref{remark1} and \ref{remark2}) following Theorem \ref{th1+}
can be drawn here concerning Theorem \ref{th1++}.

 \section{NEP With Respect to Conditional Entropy}
\label{sec2}
\setcounter{equation}{0}

Consider now an IID source pair $(X, Y) = \{(X_i, Y_i )\}_{i=1}^{\infty}$ with finite conditional entropy $H(X|Y) $, where $H(X|Y)$ is the Shannon conditional entropy of $X_i$ given $Y_i$ if $X$ is discrete, and the  conditional differential entropy of $X_i$ given $Y_i$ if $X$ is continuous. Let $p(x|y) $ be the conditional pmf or conditional pdf (as the case may be) of $X_i$ given $Y_i$, and $p(y)$ the pmf or pdf (as the case may be) of $Y_i$. By replacing $-{1 \over n} \ln p(X^n)$ with $- {1 \over n} \ln p(X^n |Y^n)$, all results and arguments in Section~\ref{sec1} can be carried over to this conditional case, yielding the NEP with respect to $H(X|Y)$.

Specifically, define
\begin{equation} \label{eq2-1}
 \lambda^* (X|Y) \defeq \sup \left\{\lambda \geq 0: \int p(y) \left [\int p^{-\lambda +1} (x|y)  d x \right ] d y < \infty \right \}
\end{equation}
where $\int d y $ is understood throughout this paper to be the summation over the source alphabet of $Y$  if $Y$ is discrete.  Suppose that
\begin{equation} \label{eq2-1+}
 \lambda^* (X|Y) >0 \;.
\end{equation}
Let
\begin{equation} \label{eq2-2}
\sigma^2_H (X|Y)  \defeq \int  \int  p(y) p(x|y) [-\ln p(x|y) ]^2 d x  d y - H^2 (X|Y)
\end{equation}
which will be referred to as the conditional information variance of $X$ given $Y$. It is not hard to see that under the assumption \eqref{eq2-1+},
\begin{equation} \label{eq2-2+}
\int \int {p(y) p^{-\lambda +1} (x|y) \over \left [ \int \int p(v)  p^{-\lambda+1} (u|v) d u d v \right ] } \left |- \ln p(x|y) \right |^k d x d y < \infty
\end{equation}
and
\[ \int \int p(y) p^{-\lambda +1} (x|y)  d x d y < \infty \]
 for any $\lambda \in (0, \lambda^* (X|Y))$ and any positive integer $k$.
Further assume that
\begin{equation} \label{eq2-1+++}
\sigma^2_H (X|Y) >0  \mbox{ and } \int \int p(y) p(x|y) | \ln p(x|y) |^3 dx d y < \infty\;.
\end{equation}
Define for any $\delta \geq 0$
\begin{equation} \label{eq2r}
 r_{X|Y} (\delta) \defeq \sup_{\lambda \geq 0} \left [ \lambda (H(X|Y) + \delta) -
 \ln \int \int p(y) p^{-\lambda +1} (x|y) d x d y \right ]
 \end{equation}
and for any $\lambda \in [0, \lambda^* (X|Y))$
 \begin{equation} \label{eq2d}
 \delta(\lambda) \defeq \int \int {p(y) p^{-\lambda +1} (x|y) \over \left [ \int \int p(v)  p^{-\lambda+1} (u|v) d u d v \right ] } \left [- \ln p(x|y) \right ]  d x dy -
  H(X|Y) \;.
\end{equation}
(Throughout this section, $\delta (\lambda)$ should be understood with its above definition.) Then under  the assumptions \eqref{eq2-1+} and \eqref{eq2-1+++}, $\delta(\lambda)$ is strictly increasing over $\lambda \in [0, \lambda^* (X|Y))$ with $\delta (0) =0$.
Let
\[ \Delta^* (X|Y) \defeq \lim_{\lambda \uparrow \lambda^* (X|Y)} \delta (\lambda) \;.\]
By an argument similar to that in the proof of Theorem~\ref{th1}, it can be shown that $r_{X|Y} (\delta)$ is strictly increasing, convex and continuously differentiable up to at least the third order inclusive over $\delta \in [0, \Delta^* (X|Y))$, and furthermore $r_{X|Y} (\delta)$ has the following parametric expression
 \begin{equation} \label{eq2p1}
  r_{X|Y} (\delta(\lambda)) =   \lambda (H(X|Y) + \delta (\lambda)) -
 \ln \int \int p(y) p^{-\lambda +1} (x|y) d x d y
 \end{equation}
with $\delta (\lambda)$ defined in \eqref{eq2d} and $\lambda = r'_{X|Y} (\delta)$.
For any $\lambda \in [0, \lambda^* (X|Y))$, define
  \begin{equation} \label{eq2-12}
  f_{\lambda} (x, y) \defeq {p^{-\lambda} (x|y) \over \int \int p(v) p^{-\lambda +1} (u|v) d u d v }
  \end{equation}
   \begin{equation} \label{eq2-13}
  \sigma^2_H (X|Y, \lambda)  \defeq \int \int f_{\lambda} (x, y) p(y) p(x|y) \left | -\ln p(x|y) - (H(X|Y) + \delta (\lambda) ) \right |^2 d x d y
  \end{equation}
\begin{equation} \label{eq2-14}
  M_H (X|Y, \lambda)  \defeq \int \int f_{\lambda} (x, y) p(y)  p(x|y)\left | -\ln p(x|y) - (H(X|Y) + \delta (\lambda) ) \right |^3 d x d y
  \end{equation}
where $\delta (\lambda)$ is defined in \eqref{eq2d}.  Write $M_H (X|Y, 0) $ as $M_H (X|Y)$. It is easy to see that $\sigma^2_H (X|Y, 0) = \sigma^2_H (X|Y)$,  $ \sigma^2_H (X|Y, \lambda) = \delta' (\lambda) $, and
\begin{equation} \label{eq2-14+}
  M_H (X|Y)  = \int \int  p(y)  p(x|y)\left | -\ln p(x|y) - H(X|Y) ) \right |^3 d x d y \;.
  \end{equation}
In parallel with Theorems~\ref{th1} and \ref{th1+}, we have the following result, which is referred to as the right NEP with respect to $H(X|Y)$ and can be proved similarly.

 \begin{theorem}[Right NEP With Respect to $H(X|Y)$] \label{th2}
For any positive integer $n$,
  \begin{equation} \label{eq2-3}
  \Pr \left \{ - {1 \over n} \ln p(X^n |Y^n ) > H(X|Y) + \delta \right \} \leq e^{- n r_{X|Y} (\delta) }
  \end{equation}
  where $X^n = X_1 X_2 \cdots X_n$ and $Y^n = Y_1 Y_2 \cdots Y_n$.
 Moreover,  under the assumptions \eqref{eq2-1+} and \eqref{eq2-1+++}, the following also hold:
 \begin{description}

 \item[(a)] There exists a $\delta^* >0$ such that for any $\delta \in (0, \delta^*]$ and any positive integer $n$,
 \begin{equation} \label{eq2-4-}
 r_{X|Y} (\delta) = {1 \over 2 \sigma^2_H (X|Y) } \delta^2 + O(\delta^3)
  \end{equation}
  and hence
 \begin{equation} \label{eq2-4}
  \Pr \left \{ - {1 \over n} \ln p( X^n|Y^n ) > H(X|Y) +\delta \right \} \leq e^{- n ({\delta^2 \over 2\sigma^2_H (X|Y) } + O(\delta^3) )}\;.
  \end{equation}

  \item[(b)] For any $\delta \in (0, \Delta^* (X|Y))$ and any positive integer $n$
  \begin{eqnarray} \label{eq2-16}
  \underline{\xi}_{H}(X|Y,\lambda,n)  e^{- n r_{X|Y} (\delta)} &\leq&
  \Pr \left \{ - {1 \over n} \ln {p(Y^n |X^n)} > H(X|Y) + \delta \right \} \nonumber\\
  &\leq& \bar{\xi}_{H}(X|Y,\lambda,n)  e^{- n r_{X|Y} (\delta)}
  \end{eqnarray}
  where $\lambda = r'_{X|Y} (\delta) >0$, and
  \begin{eqnarray}
    \label{eq2-17-1}
    \lefteqn{\bar{\xi}_{H}(X|Y,\lambda,n) = \frac{2CM_{H}(X|Y,\lambda)}{\sqrt{n} \sigma^3_{H}(X|Y,\lambda)}  } \nonumber \\
                      &&{+}\: e^{\frac{n \lambda^2 \sigma^2_{H}(X|Y,\lambda)}{2}} 
                      \left[ Q(\sqrt{n}  \lambda \sigma_{H}(X|Y,\lambda)) - Q(\rho^*+\sqrt{n}  \lambda \sigma_{H}(X|Y,\lambda))\right] \\
    \label{eq2-17-2}
   \lefteqn{ \underline{\xi}_{H}(X|Y,\lambda,n) = 
       e^{\frac{n \lambda^2 \sigma^2_{H}(X|Y,\lambda)}{2}} Q(\rho_*+\sqrt{n}  \lambda \sigma_{H}(X|Y,\lambda))}
  \end{eqnarray}
  with $Q(\rho^*) = \frac{CM_{H}(X|Y,\lambda)}{\sqrt{n} \sigma^3_{H}(X|Y,\lambda)}$ and 
  $Q(\rho_*) = \frac{1}{2} - \frac{2CM_{H}(X|Y,\lambda)}{\sqrt{n} \sigma^3_{H}(X|Y,\lambda)}$.

 \item[(c)] For any  $ \delta \leq c \sqrt{\ln n \over n} $, where $c < \sigma_H (X|Y)$ is a constant,
  \begin{eqnarray} \label{eq2-17+}
    Q  \left ( {\delta \sqrt{n} \over \sigma_H (X|Y)} \right ) - {C M_H (X|Y) \over \sqrt{n} \sigma^3_H (X|Y)}
    & \leq  &  \Pr \left \{ - {1 \over n} \ln p(X^n |Y^n ) > H(X|Y) + \delta \right \} \nonumber \\
  &  \leq &  Q  \left ( {\delta \sqrt{n} \over \sigma_H (X|Y)} \right ) + {C M_H (X|Y) \over \sqrt{n} \sigma^3_H (X|Y)} .
     \end{eqnarray}
  \end{description}

\end{theorem}

  The probability that $-{1\over n} \ln p(X^n |Y^n)$ is away from $H(X|Y)$ to the left can be bounded similarly. For completeness, we state the result without proof again. Define
\begin{equation} \label{eql2-1}
 \lambda^*_{-} (X|Y) \defeq \sup \left\{\lambda \geq 0: \int \int p(y)  p^{\lambda +1} (x|y)  d x d y < \infty \right \}\;.
\end{equation}
 Suppose that
\begin{equation} \label{eql2-1+}
 \lambda^*_{-} (X|Y) >0\;.
\end{equation}
  Define for any $\delta \geq 0$
  \[ r_{X|Y, -} (\delta) \defeq \sup_{\lambda \geq 0} \left [ \lambda (\delta- H(X|Y) ) -
 \ln \int \int p(y) p^{\lambda +1} (x|y) d x d y \right ]  \]
  and for any $\lambda \in [0, \lambda^*_{-} (X|Y))$
   \[
 \delta_{-} (\lambda) \defeq \int \int {p(y) p^{\lambda +1} (x|y) \over \left [\int \int p(v) p^{\lambda +1} (u|v) d u d v \right ] } \left [  \ln p(x|y) \right ] d x d y  + H(X|Y) \;.
 \]
(Throughout this section, $\delta_{-} (\lambda)$ should be understood with its above definition.) Then under the assumption \eqref{eq2-1+++}, $\delta_{-} (\lambda)$ is strictly increasing over $    \lambda \in [0, \lambda^*_{-} (X|Y))$ with   $\delta_{-} (0) =0$.
 Let
  \[ \Delta^*_{-} (X|Y) = \lim_{\lambda \uparrow \lambda^*_{-} (X|Y) } \delta_{-} (\lambda) \;. \]
 By using an argument similar to that in  the proof of Theorem~\ref{th1}, it can be shown that $r_{X|Y, -} (\delta)$ is strictly increasing, convex, and continuously differentiable up to at least the third order inclusive over $\delta \in [0, \Delta^*_{-} (X|Y))$, and furthermore $r_{X|Y, -} (\delta)$ has the following parametric expression
 \[ r_{X|Y, -} (\delta_{-} (\lambda)) =  \lambda (\delta_{-}(\lambda)- H(X|Y) ) -
 \ln \int \int p(y) p^{\lambda +1} (x|y) d x d y   \]
  with $\lambda = r'_{X|Y, -} (\delta) $ satisfying
    \[ \delta_{-} (\lambda) = \delta \;.\]
 Define
 \[
  \sigma^2_{H, -} (X|Y, \lambda)  \defeq \int \int {p(y) p^{\lambda +1} (x|y) \over \left [ \int \int p(v) p^{\lambda +1} (u|v) d u d v \right ] }   \left | -\ln p(x|y) - (H(X|Y) - \delta_{-} (\lambda) ) \right |^2 d x dy
 \]
and
\[
  M_{H,-} (X|Y, \lambda)  \defeq \int \int {p(y) p^{\lambda +1} (x|y) \over \left [ \int \int p(v) p^{\lambda +1} (u|v) d u d v \right ] }   \left | -\ln p(x|y) - (H(X|Y) - \delta_{-} (\lambda) ) \right |^3 d x dy \;.
\]

 In parallel with Theorem~\ref{th1++},  we have the following result, which is referred to as the left NEP with respect to $H(X|Y)$ and can be proved similarly.
 \begin{theorem}[Left NEP With Respect to $H(X|Y)$] \label{th2+}
For any positive integer $n$,
  \begin{equation} \label{eql2-3}
  \Pr \left \{ - {1 \over n} \ln p(X^n |Y^n ) \leq H(X|Y) - \delta \right \} \leq e^{- n r_{X|Y, -} (\delta) }\;.
  \end{equation}
 Furthermore,  under the assumptions \eqref{eql2-1+} and \eqref{eq2-1+++},  the following also hold:
 \begin{description}

 \item[(a)] There exists a $\delta^* >0$ such that for any $\delta \in (0, \delta^*]$ and any positive integer $n$,
 \begin{equation} \label{eql2-4-}
 r_{X|Y,-} (\delta) = {1 \over 2 \sigma^2_H (X|Y) } \delta^2 + O(\delta^3)
  \end{equation}
  and hence
 \begin{equation} \label{eql2-4}
  \Pr \left \{ - {1 \over n} \ln p( X^n |Y^n ) \leq H(X|Y) - \delta \right \} \leq e^{- n ({\delta^2 \over 2\sigma^2_H (X|Y) } + O(\delta^3) )}\;.
  \end{equation}

  \item[(b)] For any $\delta \in (0, \Delta^*_{-} (X|Y))$ and any positive integer $n$
  \begin{eqnarray} \label{eql2-16}
  \underline{\xi}_{H,-}(X|Y,\lambda,n)  e^{- n r_{X|Y,-} (\delta)} &\leq&
  \Pr \left \{ - {1 \over n} \ln {p(Y^n |X^n)} \leq H(X|Y) - \delta \right \} \nonumber\\
  &\leq& \bar{\xi}_{H,-}(X|Y,\lambda,n)  e^{- n r_{X|Y,-} (\delta)}
  \end{eqnarray}
  where $\lambda = r'_{X|Y,-} (\delta) >0$, and
  \begin{eqnarray}
    \label{eql2-17-1}
    \lefteqn{\bar{\xi}_{H,-}(X|Y,\lambda,n) = \frac{2CM_{H,-}(X|Y,\lambda)}{\sqrt{n} \sigma^3_{H,-}(X|Y,\lambda)}  } \nonumber \\
                      &&{+}\: e^{\frac{n \lambda^2 \sigma^2_{H,-}(X|Y,\lambda)}{2}} 
                      \left[ Q(\sqrt{n}  \lambda \sigma_{H,-}(X|Y,\lambda)) - Q(\rho^*+\sqrt{n}  \lambda \sigma_{H,-}(X|Y,\lambda))\right] \\
    \label{eql2-17-2}
   \lefteqn{ \underline{\xi}_{H,-}(X|Y,\lambda,n) = 
       e^{\frac{n \lambda^2 \sigma^2_{H,-}(X|Y,\lambda)}{2}} Q(\rho_*+\sqrt{n}  \lambda \sigma_{H,-}(X|Y,\lambda))}
  \end{eqnarray}
  with $Q(\rho^*) = \frac{CM_{H,-}(X|Y,\lambda)}{\sqrt{n} \sigma^3_{H,-}(X|Y,\lambda)}$ and 
  $Q(\rho_*) = \frac{1}{2} - \frac{2CM_{H,-}(X|Y,\lambda)}{\sqrt{n} \sigma^3_{H,-}(X|Y,\lambda)}$.

 \item[(c)] For any  $ \delta \leq c \sqrt{\ln n \over n} $, where $c < \sigma_H (X|Y)$ is a constant,
  \begin{eqnarray} \label{eql2-17+}
    Q  \left ( {\delta \sqrt{n} \over \sigma_H (X|Y)} \right ) - {C M_H (X|Y) \over \sqrt{n} \sigma^3_H (X|Y)}
    & \leq  &  \Pr \left \{ - {1 \over n} \ln p(X^n |Y^n ) \leq H(X|Y) - \delta \right \} \nonumber \\
  &  \leq &  Q  \left ( {\delta \sqrt{n} \over \sigma_H (X|Y)} \right ) + {C M_H (X|Y) \over \sqrt{n} \sigma^3_H (X|Y)} \;.
     \end{eqnarray}
  \end{description}
\end{theorem}

Remarks similar to those (Remark \ref{remark1} and \ref{remark2}) following Theorem \ref{th1+}
can be drawn here concerning Theorem \ref{th2} and \ref{th2+}.

Theorem~\ref{th2} will be used in \cite{jar_decoding} to show that for
any binary input memoryless channel with uniform capacity achieving
input $X$, random linear codes of block length $n$ with either Elias'
generator ensembles or Gallager's parity check ensembles can reach
within $\delta + r_{X|Y} (\delta) + {\ln n \over 2 n} -  \frac{\ln \frac{2(1-C) M_H(X|Y,\lambda)}{ \sigma^3_H (X|Y,\lambda)}  }{n}$ of the channel capacity while
maintaining word error probability upper bounded by $ (\bar{\xi}_{H}(X|Y,\lambda,n) +  \frac{2(1-C) M_H(X|Y,\lambda)}{\sqrt{n} \sigma^3_H (X|Y,\lambda)}) e^{- n r_{X|Y} (\delta)}$. In particular, when
$\delta = \sqrt{2 \alpha \sigma^2_H (X|Y)} \sqrt{\ln n \over n}$, the
word error probability is upper bounded by 
$ \frac{1}{2\sqrt{\pi \alpha \ln n}} n^{-\alpha} +O (n^{-\alpha} {\ln n \over \sqrt{n}} )$ and the
achievable rate (in nats) of  random linear codes of block length $n$
with either Elias' generator ensembles or Gallager's parity check
ensembles is within $ \sqrt{2 \alpha \sigma^2_H (X|Y)} \sqrt{\ln n
  \over n} + (\alpha+ {1\over 2}) { \ln n \over n } +O( {\ln \ln  n \over n} )$ of
the channel capacity;
when
$\delta = \frac{c}{\sqrt{n}}$
for any $c$, the word error
probability is upper bounded by  $ Q \left( \frac{c}{\sigma_H (X|Y)} \right) + {M_H (X|Y) \over
  \sigma^3_H (X|Y) \sqrt{n} }$ and the achievable rate (in nats)  is
within $ \frac{c}{\sqrt{n}} + {\ln n
  \over 2 n} -{1 \over n} \ln {(1 -C)M_H (X|Y) \over \sigma^3_H(X|Y)}
$ of the channel capacity.

We conclude this section by illustrating $r_{X|Y} (\delta)$ and $\sigma^2_H (X|Y)$ when $X$ and $Y$ are the uniform input and the corresponding output of the binary symmetric channel (BSC) and the binary input Gaussian channel.

\begin{figure}[h]
  \centering
  \includegraphics[scale=0.5]{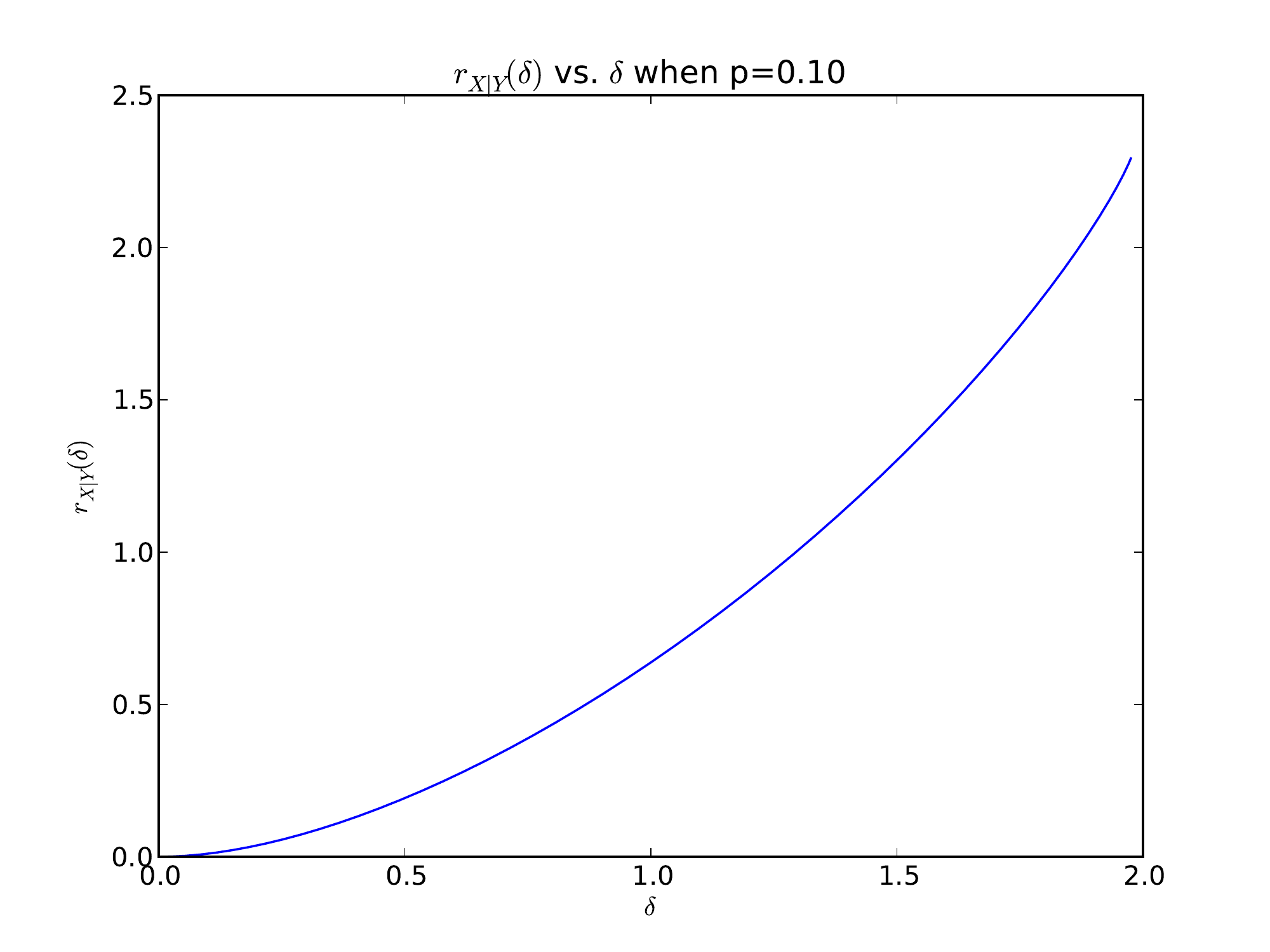}
  \caption{$r_{X|Y} (\delta)$ for BSC}
  \label{fig-bscr}
\end{figure}

{\em Example 1 (BSC)}:
Combining \eqref{eq2d} and \eqref{eq2p1}, it is not hard to verify
that
\begin{eqnarray}
  \label{eq-bsc-1}
  r_{X|Y} (\delta(\lambda)) &=& \int \int p(x,y) f_{\lambda} (x,y) \ln
  f_{\lambda} (x,y) dx dy \nonumber\\
  &=& \int \int p(x,y) f_{\lambda} (x,y) \ln
  \frac{p(x|y) f_{\lambda} (x,y)}{p(x|y)} dx dy \nonumber \\
  &\defeq& D (p(x|y)f_{\lambda}(x,y) || p(x|y))
\end{eqnarray}
For BSC, simple calculation reveals that
\begin{equation}
  \label{eq-bsc-2}
  p(x|y) = \left\{
    \begin{array}{ll}
      1 - p & \mbox{if $x=y$} \\
      p       & \mbox{otherwise}
    \end{array}
  \right.
\end{equation}
and
\begin{equation}
  \label{eq-bsc-3}
  p(x|y) f_{\lambda} (x,y) = \left\{
    \begin{array}{ll}
      \frac{(1-p)^{-\lambda+1}}{p^{-\lambda+1}+(1-p)^{-\lambda+1}} & \mbox{if $x=y$} \\
      \frac{p^{-\lambda+1}}{p^{-\lambda+1}+(1-p)^{-\lambda+1}}       & \mbox{otherwise}
    \end{array}
  \right.
\end{equation}
By defining
\begin{displaymath}
  D(q||p) \defeq (1-q) \ln \frac{1-q}{1-p} + q \ln \frac{q}{p}
\end{displaymath}
and \eqref{eq-bsc-1}, we have
\begin{eqnarray}
  \label{eq-bsc-4}
  r_{X|Y} (\delta(\lambda)) &=&
  D \left( \left. \frac{p^{-\lambda+1}}{p^{-\lambda+1}+(1-p)^{-\lambda+1}} \right\|
    p \right) \nonumber \\
  &=& D \left( \left. p +
      \frac{p(1-p)(p^{-\lambda}-(1-p)^{-\lambda})}{p^{-\lambda+1}+(1-p)^{-\lambda+1}}
      \right\|
    p \right).
\end{eqnarray}
On the other hand, by substituting \eqref{eq-bsc-2} and
\eqref{eq-bsc-3} into \eqref{eq2d},
\begin{equation}
  \label{eq-bsc-5}
  \delta (\lambda) =
  \frac{p(1-p)(p^{-\lambda}-(1-p)^{-\lambda})}{p^{-\lambda+1}+(1-p)^{-\lambda+1}}
  \ln \frac{1-p}{p}
\end{equation}
and eventually, we have
\begin{equation}
  \label{eq-bsc-6}
  r_{X|Y} (\delta) = D \left( \left. p+\frac{\delta}{\ln \frac{1-p}{p}} \right\| p \right)
\end{equation}
and plugging \eqref{eq-bsc-2} into \eqref{eq2-13} with $\lambda = 0$
yields
\begin{eqnarray}
  \label{eq-bsc-7}
  \sigma^2_{H} (X|Y) &=& (1-p) \ln^2 (1-p) + p \ln^2 p - \left[ - p \ln
    p - (1-p) \ln (1-p) \right]^2 \nonumber \\
  &=& p(1-p) \ln^2 \frac{1-p}{p}
\end{eqnarray}
Moreover, as $\mathcal{X}$ and $\mathcal{Y}$ are both finite
alphabets, it is easy to show that $\lambda^* (X|Y) = \infty$, where
$\lambda^*(X|Y)$ is defined in \eqref{eq2-1}. Then
\begin{equation}
  \label{eq-bsc-8}
  \Delta^* (X|Y) = \lim_{\lambda \uparrow +\infty} \delta
  (\lambda) = (1-p) \ln \frac{1-p}{p}
\end{equation}
and
\begin{equation}
  \label{eq-bsc-9}
  r_{\max} \defeq \lim_{\delta \uparrow  \Delta^* (X|Y)}
  r_{X|Y} (\delta) = - \ln p
\end{equation}
Based on Theorem \ref{th2}, $\Delta^*(X|Y)$ and
$r_{\max}$ can be interpreted in the following way. As
\begin{displaymath}
  \max_{x^n, y^n} - \frac{1}{n} \ln p(x^n | y^n) = - \ln p,
\end{displaymath}
then
\begin{eqnarray*}
  \lim_{\delta \to \Delta^* (X|Y)}
    \Pr \left\{ - \frac{1}{n} \ln p(X^n | Y^n) > H(X|Y) +
    \delta  \right\} &=& \Pr \left\{ - \frac{1}{n} \ln p(X^n |
    Y^n) = - \ln p
    \right\} \\
    &=& p^n = e^{ n \ln p} = e^{ - n r_{\max}}.
\end{eqnarray*}
In addition, for $\delta \geq \Delta^*(X|Y)$,
\begin{displaymath}
  \Pr \left\{ - \frac{1}{n} \ln p(X^n | Y^n) > H(X|Y) +
    \delta  \right\} = 0.
\end{displaymath}
By adopting the convention that $0 \ln 0 = 0$ and $e^{-\infty} = 0$,
\begin{equation}
  \label{eq-bsc-10}
  r_{X|Y} (\delta) =
  \left\{
    \begin{array}{ll}
      D \left( \left. p+\frac{\delta}{\ln \frac{1-p}{p}} \right\| p
      \right) & \mbox{if $\delta \in [0, \Delta^*(X|Y))$} \\
      +\infty & \mbox{if $\delta \geq \Delta^* (X|Y)$}
    \end{array}
  \right. .
\end{equation} 
A sample plot of $r_{X|Y} (\delta)$ is provided
in Figure \ref{fig-bscr} when $p=0.10$.

\begin{figure}[h]
  \centering
  \includegraphics[scale=0.5]{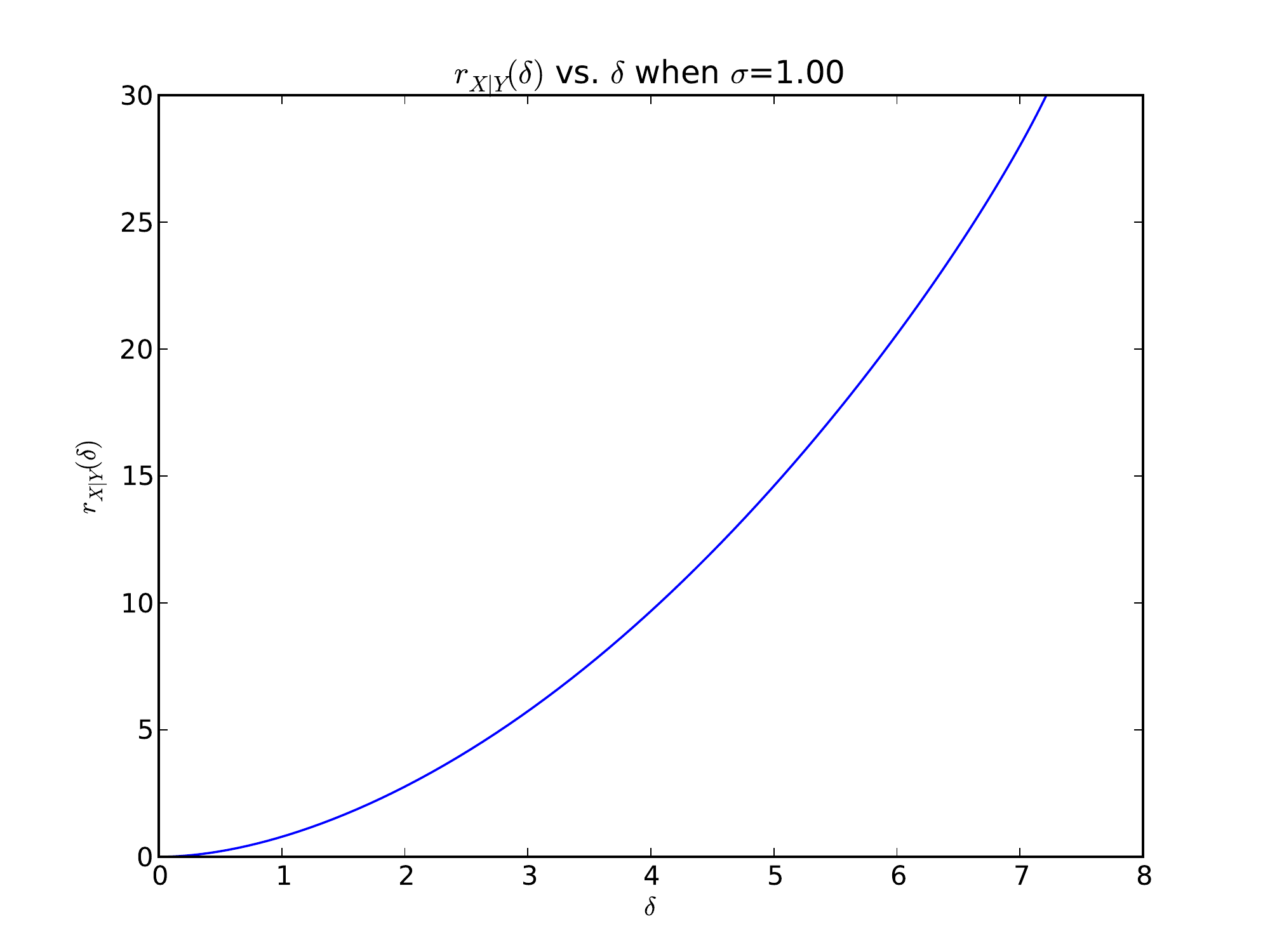}
  \caption{$r_{X|Y} (\delta)$ for BIGC}
  \label{fig-bigc}
\end{figure}

{\em Example 2 (Binary Input Gaussian Channel)}:
Without loss of generality, we assume that the input of channel is modulated to
$\{+1,-1\}$, and therefore
\begin{equation}
  \label{eq-bigc-1}
  p (y | x) = \frac{1}{ \sqrt{2 \pi} \sigma} e^{- \frac{|y-x|^2}{2 \sigma^2} }
\end{equation}
for $x = \{+1,-1\}$, where $\sigma^2$ is the variance of the noise.
Calculation of $r_{X|Y} (\delta)$ and
$\sigma^2_{H} (X|Y)$ is much more involved than that for BSC. Tedious
evaluation is omitted here with results presented as follows. Let $U$
be a standard Gaussian random variable, i.e.
\begin{displaymath}
  p(u) = \frac{1}{ \sqrt{2 \pi}} e^{- \frac{|u|^2}{2} }
\end{displaymath}
and define
\begin{displaymath}
  g (x) \defeq 1+e^{-2x}.
\end{displaymath}
Then
\begin{eqnarray}
  \label{eq-bigc-2}
  \delta (\lambda) &=&  \frac{\mathbf{E} \left[ g^{\lambda} \left(
      \frac{\sigma U+1}{\sigma^2} \right) \ln g \left(
      \frac{\sigma U+1}{\sigma^2}  \right) \right]}{ \mathbf{E} \left[ g^{\lambda} \left(
      \frac{\sigma U+1}{\sigma^2} \right) \right]} - \mathbf{E} \left[ \ln g \left(
      \frac{\sigma U+1}{\sigma^2} \right) \right] \\
  \label{eq-bigc-3}
  r_{X|Y} (\delta(\lambda)) &=&   \lambda \frac{\mathbf{E} \left[ g^{\lambda} \left(
      \frac{\sigma U+1}{\sigma^2} \right) \ln g \left(
      \frac{\sigma U+1}{\sigma^2}  \right) \right]}{ \mathbf{E} \left[ g^{\lambda} \left(
      \frac{\sigma U+1}{\sigma^2} \right) \right]} - \ln \left\{ \mathbf{E} \left[ g^{\lambda} \left(
      \frac{\sigma U+1}{\sigma^2} \right) \right] \right\}
\end{eqnarray}
and
\begin{equation}
  \label{eq-bigc-4}
  \sigma^2_{H} (X|Y) = \mathbf{E} \left[ \ln^2 g \left(
      \frac{\sigma U+1}{\sigma^2}  \right) \right] - \left\{ \mathbf{E}
    \left[ - \ln g \left(
      \frac{\sigma U+1}{\sigma^2} \right) \right] \right\}^2
\end{equation}
To get better understanding of those quantities, let us first
determine $\lambda^* (X|Y)$ and $\Delta^* (X|Y)$.
In fact, we can show that $\lambda^* (X|Y) = \infty$ by verifying that
\begin{displaymath}
  \int p(y) \left[ \sum_{x \in \mathcal{X}} p^{-\lambda + 1} (x|y) \right]
  dy < \infty
\end{displaymath}
for any finite $\lambda \geq 0$. Towards this, observe that
\begin{displaymath}
  \int p(y) \left[ \sum_{x \in \mathcal{X}} p^{-\lambda + 1} (x|y) \right]
  dy
\end{displaymath}
is an increasing function with respect to $\lambda$ since $p(x|y) \leq
1$ for any $x$ and $y$. Therefore,
\begin{eqnarray*}
  \int p(y) \left[ \sum_{x \in \mathcal{X}} p^{-\lambda + 1} (x|y) \right]
  dy &=& \mathbf{E} \left[ g^{\lambda} \left( \frac{\sigma U + 1}{\sigma^2}
    \right)\right] \\
  &\leq& \mathbf{E} \left[ g^{ \left\lceil \lambda \right\rceil} \left( \frac{\sigma U + 1}{\sigma^2}
    \right)\right] < \infty
\end{eqnarray*}
as
\begin{displaymath}
  \mathbf{E} [e^{sU}] = e^{\frac{s^2}{2}} < \infty
\end{displaymath}
for any finite $s$. Now let us show the claim $\Delta^* (X|Y) =
\infty$. According to \eqref{eq-bigc-2},
\begin{eqnarray*}
  \delta (\lambda) &=&  \frac{\mathbf{E} \left[ g^{\lambda} \left(
      \frac{\sigma U+1}{\sigma^2} \right) \ln g \left(
      \frac{\sigma U+1}{\sigma^2}  \right) \right]}{ \mathbf{E} \left[ g^{\lambda} \left(
      \frac{\sigma U+1}{\sigma^2} \right) \right]} - \mathbf{E} \left[ \ln g \left(
      \frac{\sigma U+1}{\sigma^2} \right) \right] \\
  &=& \frac{d}{d \lambda} \ln \mathbf{E} \left[ g^{\lambda} \left(
      \frac{\sigma U+1}{\sigma^2} \right) \right] - H(X|Y)
\end{eqnarray*}
As $H(X|Y)$ is a constant and always less than $\ln 2$, the claim $\Delta^* (X|Y) =
\infty$ is equivalent to show
\begin{displaymath}
  \frac{d}{d \lambda} \ln \mathbf{E} \left[ g^{\lambda} \left(
      \frac{\sigma U+1}{\sigma^2} \right) \right]
\end{displaymath}
 is unbounded when $\lambda \rightarrow \infty$. By the fact that
 $\delta (\lambda)$ is an increasing function of $\lambda$, which also
 implies that so is
\begin{displaymath}
  \frac{d}{d \lambda} \ln \mathbf{E} \left[ g^{\lambda} \left(
      \frac{\sigma U+1}{\sigma^2} \right) \right],
\end{displaymath}
we only have to verify that
\begin{displaymath}
  \frac{\ln \mathbf{E} \left[ g^{k+1} \left(
      \frac{\sigma U+1}{\sigma^2} \right) \right] - \ln \mathbf{E} \left[ g^{k} \left(
      \frac{\sigma U+1}{\sigma^2} \right) \right] }{k+1 - k} = \ln \frac{\mathbf{E} \left[ g^{k+1} \left(
      \frac{\sigma U+1}{\sigma^2} \right) \right]}{\mathbf{E} \left[ g^{k} \left(
      \frac{\sigma U+1}{\sigma^2} \right) \right]}
\end{displaymath}
or simply
\begin{displaymath}
  \frac{\mathbf{E} \left[ g^{k+1} \left(
      \frac{\sigma U+1}{\sigma^2} \right) \right]}{\mathbf{E} \left[ g^{k} \left(
      \frac{\sigma U+1}{\sigma^2} \right) \right]}
\end{displaymath}
is unbounded when $k \rightarrow \infty$, which is indeed the case as
\begin{eqnarray*}
  \frac{\mathbf{E} \left[ g^{k+1} \left(
      \frac{\sigma U+1}{\sigma^2} \right) \right]}{\mathbf{E} \left[ g^{k} \left(
      \frac{\sigma U+1}{\sigma^2} \right) \right]}  &=&
  \frac{\sum^{k+1}_{i=0}
    \left(
    \begin{array}{c}
      k+1 \\
      i
    \end{array}
    \right) e^{\frac{2i^2 - 2i}{\sigma^2}}
  }{\sum^{k}_{i=0}
    \left(
    \begin{array}{c}
      k \\
      i
    \end{array}
    \right) e^{\frac{2i^2 - 2i}{ \sigma^2}}} \\
  &=& \frac{\Theta \left( e^{\frac{2(k+1)^2 - 2(k+1)}{\sigma^2}
      }\right)}
  {\Theta \left( e^{\frac{2k^2 - 2k}{\sigma^2} }\right)} \\
  &=& \Theta \left( e^{\frac{4k}{ \sigma^2} }\right) \rightarrow \infty
\end{eqnarray*}
as $k \rightarrow \infty$. And consequently, it is not hard to see
that
\begin{displaymath}
  r_{X|Y} (\delta) \rightarrow \infty
\end{displaymath}
as $\delta \rightarrow \infty$. The interpretation based on Theorem
\ref{th2} is as follows:
\begin{displaymath}
  - \frac{1}{n} \ln p(x^n|y^n) - H(X|Y)
\end{displaymath}
can approach $\infty$ for proper choice of $x^n$ and $y^n$, but
\begin{displaymath}
  \lim_{\delta \rightarrow \infty} \Pr \left\{ - \frac{1}{n} \ln
    p(X^n|Y^n) >  H(X|Y) + \delta \right\} = e^{-\infty} = 0.
\end{displaymath}
Figure~\ref{fig-bigc} shows a sample plot of $r_{X|Y}
(\delta)$ for BIGC with $\sigma = 1.0$.

\section{NEP With Respect to Mutual Information and Relative Entropy}
\label{sec3}
\setcounter{equation}{0}

Consider now an IID source pair $(X, Y) = \{(X_i, Y_i )\}_{i=1}^{\infty}$ with finite mutual information $I(X;Y) >0$. Let $p(y|x)$ be the conditional pmf or pdf (as the case may be) of $Y_i$ given $X_i$. In this section, we extend the NEP to $I(X; Y)$ and relative entropy.

\subsection{NEP With Respect to $I(X; Y)$}

We begin with the left NEP with respect to $I(X; Y)$. Define
\begin{equation} \label{eq3-1}
 \lambda^*_{-} (X; Y) \defeq \sup \left\{\lambda \geq 0: \int \int p(x, y) \left [ {p (y|x) \over p(y) } \right ]^{-\lambda}  d x d y < \infty \right \}\;.
\end{equation}
 Suppose that
\begin{equation} \label{eq3-1+}
 \lambda^*_{-} (X; Y) >0 \;.
\end{equation}
Let
\begin{equation} \label{eq3-2}
\sigma^2_I (X;Y)  \defeq \int  \int  p(x, y) \left[ \ln { p(y|x)  \over p(y) } \right ]^2 d x  d y - I^2 (X;Y)
\end{equation}
which will be referred to as the mutual information variance of $X$ and $Y$. It is not hard to see that under the assumption \eqref{eq3-1+},
\begin{equation} \label{eq3-2+}
\int \int {p(x, y) \left [ {p (y|x) \over p(y) } \right ]^{-\lambda} \over \left [ \int \int p(u, v)  \left [ {p (v|u) \over p(v)} \right ]^{-\lambda} d u d v \right ] } \left |- \ln {p(y|x) \over p(y)}  \right |^k d x d y < \infty
\end{equation}
and
\[ \int \int  p(x, y) \left [ {p (y|x) \over p(y) } \right ]^{-\lambda} d x d y < \infty \]
 for any $\lambda \in (0, \lambda^*_{-} (X; Y))$ and any positive integer $k$.
Further assume that
\begin{equation} \label{eq3-1+++}
\sigma^2_I (X; Y) >0 \mbox{ and } \int \int p(x, y) \left | \ln {p(y|x) \over p(y) } \right |^3 d x d x <\infty . 
\end{equation}
  Define for any $\delta \geq 0$
\begin{equation} \label{eq3lr}
 r_{X;Y, -} (\delta) \defeq \sup_{\lambda \geq 0} \left [ \lambda (\delta - I(X;Y) ) -
 \ln \int \int  p(x, y) \left [ {p (y|x) \over p(y) } \right ]^{-\lambda} d x d y \right ]
 \end{equation}
and for any $\lambda \in [0, \lambda^*_{-} (X; Y))$
 \begin{equation} \label{eq3lf}
 f_{-\lambda} (x, y) \defeq  {\left [ {p (y|x) \over p(y) } \right ]^{-\lambda} \over   \int \int p(u, v)  \left [ {p (v|u) \over p(v)} \right ]^{-\lambda} d u d v  }
  \end{equation}
 \begin{equation} \label{eq3ld}
 \delta_{-} (\lambda) \defeq \int \int p(x, y) f_{-\lambda} (x, y) \left [- \ln {p(y|x) \over p(y) } \right ]  d x dy +  I(X; Y) \;.
\end{equation}
(Throughout this section, $\delta_{-} (\lambda)$ should be understood with its above definition.) Then under  the assumptions \eqref{eq3-1+} and \eqref{eq3-1+++}, $\delta_{-}(\lambda)$ is strictly increasing over $\lambda \in [0, \lambda^*_{-} (X;Y))$ with $\delta_{-} (0) =0$.
Let
\[ \Delta^*_{-} (X; Y) \defeq \lim_{\lambda \uparrow \lambda^*_{-} (X;Y)} \delta_{-} (\lambda) \;.\]
By an argument similar to that in the proof of Theorem~\ref{th1}, it can be shown that $r_{X;Y, -} (\delta)$ is strictly increasing, convex and continuously differentiable up to at least the third order inclusive over $\delta \in [0, \Delta^*_{-} (X; Y))$, and furthermore $r_{X;Y, -} (\delta)$ has the following parametric expression
 \begin{equation} \label{eq3p1}
  r_{X;Y, -} (\delta_{-}(\lambda)) =   \lambda ( \delta_{-} (\lambda) - I(X; Y)) -
  \ln \int \int  p(x, y) \left [ {p (y|x) \over p(y) } \right ]^{-\lambda} d x d y
   \end{equation}
with $\lambda = r'_{X;Y, -} (\delta)$ satisfying
 \[ \delta_{-}(\lambda) = \delta \;.  \]
Further define for any $\lambda \in [0, \lambda^*_{-} (X; Y))$
   \begin{equation} \label{eq3l-13}
  \sigma^2_{I,-} (X;Y, \lambda)  \defeq \int \int f_{-\lambda} (x, y) p(x, y) \left |
  \ln {p(y|x) \over p(y) } - (I(X; Y) - \delta_{-} (\lambda) ) \right |^2 d x d y
  \end{equation}
\begin{equation} \label{eq3l-14}
  M_{I, -} (X; Y, \lambda)  \defeq \int \int f_{-\lambda} (x, y)   p(x, y)\left | \ln {p(y|x) \over p(y) } - (I(X; Y) - \delta_{-} (\lambda) ) \right |^3 d x d y \;.
  \end{equation}
  Write $ M_{I, -} (X; Y, 0) $ simply as $M_I (X; Y)$. It is easy to see that $\sigma^2_{I, -} (X; Y, 0) = \sigma^2_I (X; Y)$,  $ \sigma^2_{I, -} (X; Y, \lambda) = \delta'_{-} (\lambda) $, and
  \begin{equation} \label{eq3l-14+}
  M_{I} (X; Y)  = \int \int    p(x, y)\left | \ln {p(y|x) \over p(y) } - I(X; Y) ) \right |^3 d x d y \;.
  \end{equation}

In parallel with Theorems~\ref{th1++} and \ref{th2+}, we have the following result, which is referred to as the left NEP with respect to $I(X; Y)$ and can be proved similarly.
\begin{theorem}[Left NEP With Respect to $I(X;Y)$] \label{th3}
For any positive integer $n$,
  \begin{equation} \label{eql3-3}
  \Pr \left \{ {1 \over n} \ln {p(Y^n |X^n)  \over p(Y^n) }\leq I(X; Y) - \delta \right \} \leq e^{- n r_{X; Y, -} (\delta) }\;.
  \end{equation}
 Furthermore,  under the assumptions \eqref{eq3-1+} and \eqref{eq3-1+++},  the following also hold:
 \begin{description}

 \item[(a)] There exists a $\delta^* >0$ such that for any $\delta \in (0, \delta^*]$ and any positive integer $n$,
 \begin{equation} \label{eql3-4-}
 r_{X; Y,-} (\delta) = {1 \over 2 \sigma^2_I (X;Y) } \delta^2 + O(\delta^3)
  \end{equation}
  and hence
 \begin{equation} \label{eql3-4}
  \Pr \left \{ {1 \over n} \ln {p(Y^n |X^n)  \over p(Y^n) } \leq I(X; Y) - \delta \right \} \leq e^{- n ({\delta^2 \over 2\sigma^2_I (X;Y) } + O(\delta^3) )}\;.
  \end{equation}

  \item[(b)] For any $\delta \in (0, \Delta^*_{-} (X; Y))$ and any positive integer $n$
 \begin{eqnarray} \label{eql3-16}
  \underline{\xi}_{I,-}(X;Y,\lambda,n)  e^{- n r_{X;Y,-} (\delta)} &\leq&
  \Pr \left \{ {1 \over n} \ln {p(Y^n |X^n)  \over p(Y^n) } \leq I(X; Y) - \delta \right\} \nonumber\\
  &\leq& \bar{\xi}_{I,-}(X;Y,\lambda,n)  e^{- n r_{X;Y,-} (\delta)}
  \end{eqnarray}
  where $\lambda = r'_{X; Y,-} (\delta) >0$, and
  \begin{eqnarray}
    \label{eql3-17-1}
    \lefteqn{\bar{\xi}_{I,-}(X;Y,\lambda,n) = \frac{2CM_{I,-}(X;Y,\lambda)}{\sqrt{n} \sigma^3_{I,-}(X;Y,\lambda)}  } \nonumber \\
                      &&{+}\: e^{\frac{n \lambda^2 \sigma^2_{I,-}(X;Y,\lambda)}{2}} 
                      \left[ Q(\sqrt{n}  \lambda \sigma_{I,-}(X;Y,\lambda)) - Q(\rho^*+\sqrt{n}  \lambda \sigma_{I,-}(X;Y,\lambda))\right] \\
    \label{eql3-17-2}
   \lefteqn{ \underline{\xi}_{I,-}(X;Y,\lambda,n) = 
       e^{\frac{n \lambda^2 \sigma^2_{I,-}(X;Y,\lambda)}{2}} Q(\rho_*+\sqrt{n}  \lambda \sigma_{I,-}(X;Y,\lambda))}
  \end{eqnarray}
  with $Q(\rho^*) = \frac{CM_{I,-}(X;Y,\lambda)}{\sqrt{n} \sigma^3_{I,-}(X;Y,\lambda)}$ and 
  $Q(\rho_*) = \frac{1}{2} - \frac{2CM_{I,-}(X;Y,\lambda)}{\sqrt{n} \sigma^3_{I,-}(X;Y,\lambda)}$.

  \item[(c)] For any  $ \delta \leq c \sqrt{\ln n \over n} $, where $c < \sigma_I (X; Y)$ is a constant,
  \begin{eqnarray} \label{eql3-17+}
    Q  \left ( {\delta \sqrt{n} \over \sigma_I (X; Y)} \right ) - {C M_I (X; Y) \over \sqrt{n} \sigma^3_I (X; Y)}
    & \leq  &
    \Pr \left \{ {1 \over n} \ln {p(Y^n |X^n)  \over p(Y^n) }\leq I(X; Y) - \delta \right \}
    \nonumber \\
  &  \leq &  Q  \left ( {\delta \sqrt{n} \over \sigma_I (X; Y)} \right ) + {C M_I (X; Y) \over \sqrt{n} \sigma^3_I (X; Y)}.
     \end{eqnarray}
  \end{description}
\end{theorem}

The probability that ${1 \over n} \ln {p(Y^n |X^n)  \over p(Y^n) } $ is away from $I(X; Y)$ to the right can be bounded in a similar manner. For completeness, we state these bounds again without proof. Define
\begin{equation} \label{eqr3-1}
 \lambda^* (X; Y) \defeq \sup \left\{\lambda \geq 0: \int \int p(x, y) \left [ {p (y|x) \over p(y) } \right ]^{\lambda}  d x d y < \infty \right \}\;.
\end{equation}
 Suppose that
\begin{equation} \label{eqr3-1+}
 \lambda^*(X; Y) >0 \;.
\end{equation}
Define for any $\delta \geq 0$
\begin{equation} \label{eq3r}
 r_{X;Y} (\delta) \defeq \sup_{\lambda \geq 0} \left [ \lambda (I(X;Y) + \delta) -
 \ln \int \int  p(x, y) \left [ {p (y|x) \over p(y) } \right ]^{\lambda} d x d y \right ]
 \end{equation}
and for any $\lambda \in [0, \lambda^*(X; Y))$
 \begin{equation} \label{eq3f}
 f_{\lambda} (x, y) \defeq  {\left [ {p (y|x) \over p(y) } \right ]^{\lambda} \over   \int \int p(u, v)  \left [ {p (v|u) \over p(v)} \right ]^{\lambda} d u d v  }
  \end{equation}
 \begin{equation} \label{eq3d}
 \delta (\lambda) \defeq \int \int p(x, y) f_{\lambda} (x, y) \left [ \ln {p(y|x) \over p(y) } \right ]  d x dy -  I(X; Y) \;.
\end{equation}
(Throughout this section, $\delta (\lambda)$ should be understood with its above definition.) Then under  the assumptions \eqref{eqr3-1+} and \eqref{eq3-1+++}, $\delta(\lambda)$ is strictly increasing over $\lambda \in [0, \lambda^* (X;Y))$ with $\delta(0) =0$.
Let
\[ \Delta^* (X; Y) \defeq \lim_{\lambda \uparrow \lambda^* (X;Y)} \delta (\lambda) \;.\]
By an argument similar to that in the proof of Theorem~\ref{th1}, it can be shown that $r_{X;Y} (\delta)$ is strictly increasing, convex and continuously differentiable up to at least the third order over $\delta \in [0, \Delta^* (X; Y))$, and furthermore $r_{X;Y} (\delta)$ has the following parametric expression
 \begin{equation} \label{eq3p}
  r_{X;Y} (\delta(\lambda)) =   \lambda ( I (X; Y) + \delta(\lambda) ) -
  \ln \int \int  p(x, y) \left [ {p (y|x) \over p(y) } \right ]^{\lambda} d x d y
   \end{equation}
with $\lambda = r'_{X;Y} (\delta)$ satisfying
 \[ \delta (\lambda) = \delta \;.  \]
Further define for any $\lambda \in [0, \lambda^*(X; Y))$
   \begin{equation} \label{eq3-13}
  \sigma^2_{I} (X;Y, \lambda)  \defeq \int \int f_{\lambda} (x, y) p(x, y) \left |
  \ln {p(y|x) \over p(y) } - (I(X; Y) + \delta (\lambda) ) \right |^2 d x d y
  \end{equation}
\begin{equation} \label{eq3-14}
  M_{I} (X; Y, \lambda)  \defeq \int \int f_{\lambda} (x, y)   p(x, y)\left | \ln {p(y|x) \over p(y) } - (I(X; Y) +\delta (\lambda) ) \right |^3 d x d y \;.
  \end{equation}
  It is easy to see that $\sigma^2_{I} (X; Y, 0) = \sigma^2_I (X; Y)$ and $ \sigma^2_{I} (X; Y, \lambda) = \delta' (\lambda) $.

In parallel with Theorems~\ref{th1}, \ref{th1+},  and \ref{th2}, we
have the following result, which is referred to as the right NEP with
respect to $I(X; Y)$ and can be proved similarly.
\begin{theorem}[Right NEP With Respect to $I(X;Y)$] \label{th3+}
For any positive integer $n$,
  \begin{equation} \label{eq3-3}
  \Pr \left \{ {1 \over n} \ln {p(Y^n |X^n)  \over p(Y^n) } > I(X; Y) + \delta \right \} \leq e^{- n r_{X; Y} (\delta) }\;.
  \end{equation}
 Furthermore,  under the assumptions \eqref{eqr3-1+} and \eqref{eq3-1+++},  the following also hold:
 \begin{description}

 \item[(a)] There exists a $\delta^* >0$ such that for any $\delta \in (0, \delta^*]$ and any positive integer $n$,
 \begin{equation} \label{eq3-4-}
 r_{X; Y} (\delta) = {1 \over 2 \sigma^2_I (X;Y) } \delta^2 + O(\delta^3)
  \end{equation}
  and hence
 \begin{equation} \label{eq3-4}
  \Pr \left \{ {1 \over n} \ln {p(Y^n |X^n)  \over p(Y^n) } > I(X; Y) + \delta \right \} \leq e^{- n ({\delta^2 \over 2\sigma^2_I (X;Y) } + O(\delta^3) )}\;.
  \end{equation}

  \item[(b)] For any $\delta \in (0, \Delta^* (X; Y))$ and any positive integer $n$
  \begin{eqnarray} \label{eq3-16}
  \underline{\xi}_{I}(X;Y,\lambda,n)  e^{- n r_{X;Y} (\delta)} &\leq&
  \Pr \left \{  {1 \over n} \ln {p(Y^n |X^n)  \over p(Y^n) } > I(X; Y) + \delta  \right \} \nonumber\\
  &\leq& \bar{\xi}_{I}(X;Y,\lambda,n)  e^{- n r_{X;Y} (\delta)}
  \end{eqnarray}
  where $\lambda = r'_{X; Y} (\delta) >0$, and 
  \begin{eqnarray}
    \label{eq3-17-1}
    \lefteqn{\bar{\xi}_{I}(X;Y,\lambda,n) = \frac{2CM_{I}(X;Y,\lambda)}{\sqrt{n} \sigma^3_{I}(X;Y,\lambda)}  } \nonumber \\
                      &&{+}\: e^{\frac{n \lambda^2 \sigma^2_{I}(X;Y,\lambda)}{2}} 
                      \left[ Q(\sqrt{n}  \lambda \sigma_{I}(X;Y,\lambda)) - Q(\rho^*+\sqrt{n}  \lambda \sigma_{I}(X;Y,\lambda))\right] \\
    \label{eq3-17-2}
   \lefteqn{ \underline{\xi}_{I}(X;Y,\lambda,n) = 
       e^{\frac{n \lambda^2 \sigma^2_{I}(X;Y,\lambda)}{2}} Q(\rho_*+\sqrt{n}  \lambda \sigma_{I}(X;Y,\lambda))}
  \end{eqnarray}
  with $Q(\rho^*) = \frac{CM_{I}(X;Y,\lambda)}{\sqrt{n} \sigma^3_I(X;Y,\lambda)}$ and 
  $Q(\rho_*) = \frac{1}{2} - \frac{2CM_I(X;Y,\lambda)}{\sqrt{n} \sigma^3_I(X;Y,\lambda)}$.
  \item[(c)] For any  $ \delta \leq c \sqrt{\ln n \over n} $, where $c < \sigma_I (X; Y)$ is a constant,
  \begin{eqnarray} \label{eq3-17+}
    Q  \left ( {\delta \sqrt{n} \over \sigma_I (X; Y)} \right ) - {C M_I (X; Y) \over \sqrt{n} \sigma^3_I (X; Y)}
    & \leq  &
    \Pr \left \{ {1 \over n} \ln {p(Y^n |X^n)  \over p(Y^n) } > I(X; Y) + \delta \right \}
    \nonumber \\
  &  \leq &  Q  \left ( {\delta \sqrt{n} \over \sigma_I (X; Y)} \right ) + {C M_I (X; Y) \over \sqrt{n} \sigma^3_I (X; Y)} \;.
     \end{eqnarray}
  \end{description}
\end{theorem}

Remarks similar to those (Remark \ref{remark1} and \ref{remark2}) following Theorem \ref{th1+}
can be drawn here concerning Theorems \ref{th3} and \ref{th3+}.

\subsection{NEP With Respect to Relative Entropy}

The IID source pair $(X, Y) = \{ (X_i, Y_i ) \}_{i=1}^{\infty}$ considered so far is arbitrary. Let us now focus on the case in which the source $X$ is discrete, but $Y$ could be either discrete or continuous. Let ${\cal P}$ denote the set of all probability distributions over the source alphabet $\cal X$. For any $t \in {\cal P}$, let
 \begin{equation} \label{eq3r-1}
 q_t (y) \defeq \sum_{x \in {\cal X}} t(x) p(y |x)
 \end{equation}
 \begin{equation} \label{eq3r-2}
 q_t (y^n) \defeq \prod_{i=1}^n q_t (y_i)
 \end{equation}
 \begin{equation} \label{eq3r-2+}
 D(t, x) \defeq \int p(y |x) \ln {p(y|x) \over q_t (y) } d y
  \end{equation}
 and
 \begin{equation} \label{eq3r-3}
 I (t; P) \defeq \sum_{x \in {\cal X}} t(x) \int p(y |x) \ln {p(y|x) \over q_t (y) } d y
 \end{equation}
 where $y^n = y_1 y_2 \cdots y_n$, and $P = \{ p(y |x ) \}$ represents a channel with $p(y|x)$ as its transitional pmf or pdf (as the case may be). Clearly, $D(t, x)$ is the divergence or relative entropy between $p(y|x)$ and $q_t (y)$;  and $I(t; P)$ is the mutual information between the input and output of the channel $P $ when the input is distributed according to $t$. To be specific, we denote the pmf of each $X_i$ by $p_X$. Without loss of generality, we assume that $p_X (x) >0$ for any $x \in {\cal X}$. Since
  \[
  \int \int p(x, y) \left [ {p (y|x) \over p(y) } \right ]^{-\lambda}  d x d y
   =  \sum_{a \in {\cal X}} p_X (a) \int p(y |a) \left [ {\sum_{b \in {\cal X}} p_X (b) p(y |b)  \over p(y |a) } \right ]^{\lambda} d y
 \]
 it is not hard to see that for any $\lambda >0$,
   \[
  \int \int p(x, y) \left [ {p (y|x) \over p(y) } \right ]^{-\lambda}  d x d y < \infty \]
  if and only if
 \[   \int p(y |a) \left [ {\sum_{b \in {\cal X}}  p(y |b)  \over p(y |a) } \right ]^{\lambda} d y  <\infty
 \]
for any $a \in {\cal X}$. Therefore, $\lambda^*_{-} (X; Y)$ defined in \eqref{eq3-1} is also equal to
  \[ \sup \left \{ \lambda \geq 0:  \int p(y |a) \left [ {p(y|a)  \over q_t (y) } \right ]^{-\lambda} d y  <\infty, a \in {\cal X} \right \} \]
  for any $t \in {\cal P}$ with $t(a) >0 $ for any $a \in {\cal X}$ (such $t \in {\cal P}$ will be said to have full support).

 Define for any $t \in {\cal P}$ with full support and any $\delta \geq 0$
\begin{equation} \label{eq3rlr}
 r_{-} (t, \delta) \defeq \sup_{\lambda \geq 0} \left [ \lambda (\delta - I(t; P) ) -
 \sum_{x \in {\cal X}} t(x) \ln  \int  p(y |x) \left [ {p (y|x) \over q_t (y) } \right ]^{-\lambda}  d y \right ]
 \end{equation}
and for any $\lambda \in [0, \lambda^*_{-} (X; Y))$ and any $t \in {\cal P}$ with full support
 \begin{equation} \label{eq3rlf}
 f_{-\lambda} (y |x) \defeq  {\left [ {p (y|x) \over q_t (y) } \right ]^{-\lambda} \over    \int p(v |x )  \left [ {p (v|x ) \over q_t (v)} \right ]^{-\lambda} d v  }
  \end{equation}
 \begin{equation} \label{eq3rld-}
 D(t, x, \lambda) \defeq \int p( y|x) f_{-\lambda} (y |x) \left [\ln {p(y|x) \over  q_t(y) } \right ]   dy
 \end{equation}
 \begin{equation} \label{eq3rld}
 \delta_{-} (t, \lambda) \defeq \sum_{x \in {\cal X}} t (x)  \int p( y|x) f_{-\lambda} (y |x) \left [- \ln {p(y|x) \over  q_t(y) } \right ]   dy +  I(t; P) \;.
\end{equation}
It is not hard to verify that
 \[  \delta_{-} (t, 0)  =0 \]
 and
 \begin{eqnarray*}
  {\partial  \delta_{-} (t, \lambda)  \over \partial \lambda}
  & = &
  \sum_{x \in {\cal X}} t (x) \left [ \int p( y|x) f_{-\lambda} (y |x) \left [- \ln {p(y|x) \over  q_t(y) } \right ]^2   dy   \right. \\
  & & \left.- \left ( \int p( y|x) f_{-\lambda} (y |x) \left [- \ln {p(y|x) \over  q_t(y) } \right ]   dy \right )^2  \right] \\
 & =&   \sum_{x \in {\cal X}} t (x) \left [ \int p( y|x) f_{-\lambda} (y |x) \left [ \ln {p(y|x) \over  q_t(y) } \right ]^2   dy   -  D^2 (t, x, \lambda) \right ] \\
 & > & 0
  \end{eqnarray*}
  where the last inequality is due to \eqref{eq3-1+++}.
  Therefore,   $\delta_{-}(t, \lambda)$ as a function of $\lambda$ is strictly increasing over $\lambda \in [0, \lambda^*_{-} (X;Y))$.
Let
\[ \Delta^*_{-} (t) \defeq \lim_{\lambda \uparrow \lambda^*_{-} (X;Y)} \delta_{-} (t, \lambda) \;.\]
By an argument similar to that in the proof of Theorem~\ref{th1}, it can be shown that $r_{ -} (t, \delta)$ is strictly increasing, convex and continuously differentiable up to at least the third  order inclusive over $\delta \in [0, \Delta^*_{-} (t))$, and furthermore $r_{-} (t, \delta)$ has the following parametric expression
 \begin{equation} \label{eq3rp1}
  r_{ -} (t, \delta_{-}(t, \lambda)) =   \lambda ( \delta_{-} (t, \lambda) - I(t; P)) -
  \sum_{x \in {\cal X}} t(x) \ln  \int  p(y |x) \left [ {p (y|x) \over q_t (y) } \right ]^{-\lambda}  d y
   \end{equation}
with
\[ \lambda = {\partial r_{-} (t, \delta) \over \partial \delta }\]
 satisfying
 \[ \delta_{-}(t, \lambda) = \delta \;.  \]
Further define for any $\lambda \in [0, \lambda^*_{-} (X; Y))$
   \begin{equation} \label{eq3rl-13}
  \sigma^2_{D, -} (t; P, \lambda)  \defeq \sum_{x \in {\cal X}} t (x) \left [ \int p( y|x) f_{-\lambda} (y |x)   \left |
  \ln {p(y|x) \over q_t (y) } - D(t, x, \lambda) \right |^2  d y \right ]
  \end{equation}
  and
\begin{equation} \label{eq3rl-14}
  M_{D, -} (t; P , \lambda)  \defeq \sum_{x \in {\cal X}} t (x) \left [ \int p( y|x) f_{-\lambda} (y |x)   \left |
  \ln {p(y|x) \over q_t (y) } - D(t, x, \lambda)  \right |^3  d y \right ]
  \;.
  \end{equation}
  Write $\sigma^2_{D, -} (t; P, 0)  $ simply as $\sigma^2_D (t; P)$, $M_{D, -} (t; P , 0)$ as $M_{D} (t; P )$, $\sigma^2_D (p_X; P)$ as $\sigma^2_D (X; Y)$, and $M_{D} (p_X; P )$ as $M_D (X; Y)$. It is not hard to see that
  \[ \sigma^2_D (t; P) = \sum_{x \in {\cal X}} t (x) \left [ \int p( y|x)  \left |
  \ln {p(y|x) \over q_t (y) }  \right |^2  d y - \left ( \int p( y|x)
  \ln {p(y|x) \over q_t (y) }    d y \right )^2  \right ] \]
 \[ \sigma^2_D (X; Y) = \sum_{x \in {\cal X}} p(x) \left [ \int p( y|x)  \left |
  \ln {p(y|x) \over p (y) }  \right |^2  d y - \left ( \int p( y|x)
  \ln {p(y|x) \over p (y) }    d y \right )^2  \right ] \]
  \[ M_{D} (t; P)  \defeq \sum_{x \in {\cal X}} t (x) \left [ \int p( y|x)  \left |
  \ln {p(y|x) \over q_t (y) } -  \left ( \int p( v|x)     \ln {p(v|x) \over q_t (v) } dv \right )
 \right |^3  d y \right ] \]
  \[ M_{D} (X; Y)  \defeq \sum_{x \in {\cal X}} p (x) \left [ \int p( y|x)  \left |
  \ln {p(y|x) \over p (y) } -  \left ( \int p( v|x)     \ln {p(v|x) \over p (v) } dv \right )
 \right |^3  d y \right ] \] and
 \[ \sigma^2_{D, -} (t; P, \lambda)  =  {\partial  \delta_{-} (t, \lambda)  \over \partial \lambda}\;. \]
 For obvious reasons, we will refer to $\sigma^2_D (t; P)$ ($\sigma^2_D (X; Y)$, respectively)
 as the conditional divergence (or relative entropy) variance of $P$ given $t$ ($Y$ given $X$, respectively).

In parallel with Theorems~\ref{th1++}, \ref{th2+}, and \ref{th3},  we have the following result, which is referred to as the left NEP with respect to relative entropy.
\begin{theorem}[Left NEP With Respect to Relative Entropy] \label{th3r}
For any sequence $x^n=x_1 \cdots x_n$ from $\cal X$, let $t \in {\cal P}$ be the type of $x^n$, i.e., $n t(a)$, $a \in {\cal X}$,  is the number of times the symbol $a$ appears in $x^n$. Assume that $t$ has full support. Then
  \begin{equation} \label{eqrl3-3}
  \Pr \left \{ \left. {1 \over n} \ln {p(Y^n |X^n)  \over q_t (Y^n) }\leq I(t; P) - \delta \right |  X^n =x^n    \right \} \leq e^{- n r_{-} (t, \delta) }\;.
  \end{equation}
 Furthermore,  under the assumptions \eqref{eq3-1+} and \eqref{eq3-1+++},  the following also hold:
 \begin{description}

 \item[(a)] There exists a $\delta^* >0$ such that for any $\delta \in (0, \delta^*]$  \begin{equation} \label{eqrl3-4-}
 r_{-} (t, \delta) = {1 \over 2 \sigma^2_D (t; P) } \delta^2 + O(\delta^3)
  \end{equation}
  and hence
 \begin{equation} \label{eqrl3-4}
  \Pr \left \{ \left. {1 \over n} \ln {p(Y^n |X^n)  \over q_t(Y^n) }\leq I(t; P) - \delta \right | X^n = x^n  \right \} \leq e^{- n ({\delta^2 \over 2\sigma^2_D (t; P) } + O(\delta^3) )}\;.
  \end{equation}

  \item[(b)] For any $\delta \in (0, \Delta^*_{-} (X; Y))$
  \begin{eqnarray} \label{eqrl3-17}
  \underline{\xi}_{D,-}(t;P,\lambda,n)  e^{- n r_{-} (t, \delta)} &\leq&
  \Pr \left \{ \left. {1 \over n} \ln {p(Y^n |X^n)  \over q_t(Y^n) }\leq I(t; P) - \delta \right | X^n = x^n  \right \} \nonumber\\
  &\leq& \bar{\xi}_{D,-}(t;P,\lambda,n)  e^{- n r_{-} (t, \delta)}
  \end{eqnarray}
  where $\lambda = {\partial  r_{-} (t, \delta)  \over \partial \delta} >0$, and
  \begin{eqnarray}
    \label{eqrl3-17-1}
    \lefteqn{\bar{\xi}_{D,-}(t;P,\lambda,n) = \frac{2CM_{D,-}(t;P,\lambda)}{\sqrt{n} \sigma^3_{D,-}(t;P,\lambda)}  } \nonumber \\
                      &&{+}\: e^{\frac{n \lambda^2 \sigma^2_{D,-}(t;P,\lambda)}{2}} 
                      \left[ Q(\sqrt{n}  \lambda \sigma_{D,-}(t;P,\lambda)) - Q(\rho^*+\sqrt{n}  \lambda \sigma_{D,-}(t;P,\lambda))\right] \\
    \label{eqrl3-17-2}
    \lefteqn{ \underline{\xi}_{D,-}(t;P,\lambda,n) = 
       e^{\frac{n \lambda^2 \sigma^2_{D,-}(t;P,\lambda)}{2}} Q(\rho_*+\sqrt{n}  \lambda \sigma_{D,-}(t;P,\lambda))}
  \end{eqnarray}
  with $Q(\rho^*) = \frac{CM_{D,-}(t;P,\lambda)}{\sqrt{n} \sigma^3_{D,-}(t;P,\lambda)}$ and 
  $Q(\rho_*) = \frac{1}{2} - \frac{2CM_{D,-}(t;P,\lambda)}{\sqrt{n} \sigma^3_{D,-}(t;P,\lambda)}$.
  
  \item[(c)] For any  $ \delta \leq c \sqrt{\ln n \over n} $, where $c < \sigma_D (t; P)$ is a constant,
  \begin{eqnarray} \label{eqrl3-17+}
    Q  \left ( {\delta \sqrt{n} \over \sigma_D (t; P)} \right ) - {C M_D (t; P) \over \sqrt{n} \sigma^3_D (t; P)}
    & \leq  &   \Pr \left \{ \left. {1 \over n} \ln {p(Y^n |X^n)  \over q_t(Y^n) }\leq I(t; P) - \delta \right | X^n = x^n  \right \}
    \nonumber \\
  &  \leq &  Q  \left ( {\delta \sqrt{n} \over \sigma_D (t; P)} \right ) + {C M_D (t; P) \over \sqrt{n} \sigma^3_D (t; P)} .
     \end{eqnarray}
  \end{description}
\end{theorem}

  \begin{IEEEproof}[Proof of Theorem~\ref{th3r}]
   The inequality \eqref{eqrl3-3} comes from the Chernoff bound. To see this is indeed the case, note that
  \begin{eqnarray} \label{eq3-d1}
  \lefteqn{ \Pr \left \{ \left. {1 \over n} \ln {p(Y^n |X^n)  \over q_t (Y^n) }\leq I(t; P) - \delta \right |  X^n =x^n    \right \}} \nonumber \\
  & \leq & \inf_{\lambda \geq 0} { \be \left [ \left (  \left.  {p(Y^n |X^n)  \over q_t (Y^n) } \right )^{-\lambda} \right | X^n = x^n \right ] \over e^{n \lambda ( \delta - I(t; P))} } \nonumber \\
  & = & \inf_{\lambda \geq 0} { \prod_{a \in {\cal X}}  \left [ \int p(y|a )  \left ({p(y |a)  \over q_t (y) } \right )^{-\lambda}  d y  \right ] ^{n t(a)} \over e^{n \lambda ( \delta - I(t; P))} } \nonumber \\
  & = & \inf_{\lambda \geq 0} \exp \left \{ -n \left [  \lambda ( \delta - I(t; P)) - \sum_{a \in {\cal X}} t(a) \ln   \int p(y|a )  \left ({p(y |a)  \over q_t (y) } \right )^{-\lambda}  d y  \right ] \right \} \nonumber \\
  & = & e^{-n r_{-} (t, \delta) }
  \end{eqnarray}
  which completes the proof of \eqref{eqrl3-3}.

  The equation \eqref{eqrl3-4-} follows from the Taylor expansion of $r_{-} (t, \delta)$ at $\delta =0$ and the fact that
  \[   {\partial^2  r_{-} (t, \delta)  \over \partial \delta^2}   = { 1 \over \sigma^2_D (t; P) }\;.\]
 What remains is to prove \eqref{eqrl3-17} and \eqref{eqrl3-17+}. To this end, let
 \[ f_{-\lambda} (y^n |x^n) = \prod_{i=1}^n f_{-\lambda} (y_i | x_i ). \]
 With $\lambda = {\partial  r_{-} (t, \delta)  \over \partial \delta}$, it follows from
 \eqref{eq3rp1} that
 \[  r_{ -} (t, \delta ) =   \lambda ( \delta - I(t; P)) -
  \sum_{x \in {\cal X}} t(x) \ln  \int  p(y |x) \left [ {p (y|x) \over q_t (y) } \right ]^{-\lambda}  d y  \;.\]
  Then we have
  \begin{eqnarray} \label{eq3-d2}
  \lefteqn{\Pr \left \{ \left. {1 \over n} \ln {p(Y^n |X^n)  \over q_t(Y^n) }\leq I(t; P) - \delta \right | X^n =x^n  \right \}} \nonumber \\
  & = & \int\limits_{{1 \over n} \ln {p(y^n |x^n)  \over q_t(y^n) }\leq I(t; P) - \delta} p(y^n | x^n) d y^n \nonumber \\
  & = & \int\limits_{{1 \over n} \ln {p(y^n |x^n)  \over q_t(y^n) }\leq I(t; P) - \delta}
            f^{-1}_{-\lambda} (y^n |x^n) f_{-\lambda} (y^n | x^n)  p(y^n | x^n) d y^n \nonumber \\
  & = & \int\limits_{{1 \over n} \ln {p(y^n |x^n)  \over q_t(y^n) }\leq I(t; P) - \delta}
            e^{ \lambda \ln {p(y^n |x^n)  \over q_t (y^n)} +
                  n \sum_{a \in {\cal X}} t(a) \ln   \int p(v|a )  \left ({p(v |a)  \over q_t (v) } \right )^{-\lambda}  d v  }   
            f_{-\lambda} (y^n | x^n)  p(y^n | x^n) d y^n    \nonumber \\
  & = & \int\limits_{{1 \over n} \ln {p(y^n |x^n)  \over q_t(y^n) }\leq I(t; P) - \delta}
            e^{ \lambda \ln {p(y^n |x^n)  \over q_t (y^n)} +
                  n \lambda ( \delta - I(t; P)) - n r_{ -} (t, \delta )   }   
            f_{-\lambda} (y^n | x^n)  p(y^n | x^n) d y^n    \nonumber \\
  & = & e^{-n r_{ -} (t, \delta )} \int\limits_{ \ln {p(y^n |x^n)  \over q_t(y^n) } - n(I(t; P) - \delta)  \leq 0}
            e^{ \lambda \left[ \ln {p(y^n |x^n)  \over q_t (y^n)} - n(I(t; P) - \delta) \right]   }   
            f_{-\lambda} (y^n | x^n)  p(y^n | x^n) d y^n    \nonumber \\
  & = & e^{-n r_{ -} (t, \delta )} \int\limits_{\rho \leq 0}
            \int\limits_{ \frac{\ln {p(y^n |x^n)  \over q_t(y^n) }  - n(I(t; P) - \delta)}{\sqrt{n} \sigma_{D,-} (t;P,\lambda)} = \rho}
            e^{ \lambda \sqrt{n} \sigma_{D,-} (t;P,\lambda) \rho  }   
            f_{-\lambda} (y^n | x^n)  p(y^n | x^n) d y^n    \nonumber \\
  & = & e^{-n r_{ -} (t, \delta )} \int\limits^{0}_{-\infty} 
            e^{ \lambda \sqrt{n} \sigma_{D,-} (t;P,\lambda) \rho  } d F_{x^n} (\rho) \nonumber \\
  & = & e^{-n r_{ -} (t, \delta )} \left[ F_{x^n}(0) - \int\limits^{0}_{-\infty} \lambda \sqrt{n} \sigma_{D,-} (t;P,\lambda)
            e^{ \lambda \sqrt{n} \sigma_{D,-} (t;P,\lambda) \rho  } F_{x^n} (\rho) d \rho \right] \;.
   \end{eqnarray}
  where
  \begin{displaymath}
    F_{x^n} (\rho) =
    \Pr \left\{ \frac{\ln {p(Z^n |x^n)  \over q_t(Z^n) }  - n(I(t; P) - \delta)  }{\sqrt{n} \sigma_{D,-} (t;P,\lambda)} \leq \rho 
                   \right\}
  \end{displaymath}
  and $Z_i$ takes values over the alphabet of $Y$ according to the pmf or pdf (as the case may be) $f_{-\lambda} (z | x_i)  p(z | x_i)  $.
  It is easy to verify that
  \[ \be \left [ \ln {p(Z_i |x_i)  \over q_t (Z_i) }  \right ] = D (t, x_i, \lambda) \]
  and
  \begin{eqnarray*}
    \sum_{i =1}^n \be \left [ \ln {p(Z_i |x_i)  \over q_t (Z_i) }  \right ]
  & = & \sum^n_{i=1} D(t,x_i,\lambda) \\
 &  =  & n \sum_{x \in {\cal X}} t(x) D(t, x, \lambda) \\
 & = & n(I(t; P) - \delta) 
 \end{eqnarray*}
 which further implies that
 \begin{displaymath}
   F_{x^n} (\rho) =
    \Pr \left\{ \frac{ \sum^n_{i=1} \left[ \ln {p(Z_i |x_i)  \over q_t(Z_i) } - D(t,x_i,\lambda) \right]  }
                 {\sqrt{n} \sigma_{D,-} (t;P,\lambda)} \leq \rho \right\} \;.
 \end{displaymath}
 Applying Lemma~\ref{le1} to the independent sequence
  \[ \left \{ \ln {p(Z_i |x_i)  \over q_t (Z_i) } - D (t, x_i, \lambda)  \right \}_{i=1}^n , \]
 the argument similar to that in the proof of Theorem~\ref{th1+} can then be used to establish \eqref{eqrl3-17}.

  Finally, consider another sequence of independent random variables $W_1, W_2, \cdots, W_n$, where $W_i$ takes values over the alphabet of $Y$ according to the pmf or pdf (as the case may be) $ p(w | x_i)  $.  Applying Lemma~\ref{le1} directly to
  \[ \left \{ \ln {p(W_i |x_i)  \over q_t (W_i) } - D (t, x_i)  \right \}_{i=1}^n \]
we then get   \eqref{eqrl3-17+}.  This completes the proof of Theorem~\ref{th3r}.
\end{IEEEproof}

  The conditional probability that given $X^n = x^n$, ${1 \over n} \ln {p(Y^n |X^n)  \over q_t(Y^n) }   $ is away from $I(t; P)$ to the right can be bounded similarly. For completeness, we state these bounds below without proof. Define for any $t \in {\cal P}$ with full support and any $\delta \geq 0$
\begin{equation} \label{eq3rr}
 r (t, \delta) \defeq \sup_{\lambda \geq 0} \left [ \lambda (I(t; P) + \delta) -
 \sum_{x \in {\cal X}} t(x) \ln  \int  p(y |x) \left [ {p (y|x) \over q_t (y) } \right ]^{\lambda}  d y \right ]
 \end{equation}
and for any $\lambda \in [0, \lambda^* (X; Y))$ and any $t \in {\cal P}$ with full support
 \begin{equation} \label{eq3rf}
 f_{\lambda} (y |x) \defeq  {\left [ {p (y|x) \over q_t (y) } \right ]^{\lambda} \over    \int p(v |x )  \left [ {p (v|x ) \over q_t (v)} \right ]^{\lambda} d v  }
  \end{equation}
 \begin{equation} \label{eq3rd-}
 D_{+}(t, x, \lambda) \defeq \int p( y|x) f_{\lambda} (y |x) \left [\ln {p(y|x) \over  q_t(y) } \right ]   dy
 \end{equation}
 \begin{equation} \label{eq3rd}
 \delta (t, \lambda) \defeq \sum_{x \in {\cal X}} t (x)  \int p( y|x) f_{\lambda} (y |x) \left [\ln {p(y|x) \over  q_t(y) } \right ]   dy -  I(t; P) \;.
\end{equation}
Then under the condition \eqref{eq3-1+++},   $\delta(t, \lambda)$ as a function of $\lambda$ is strictly increasing over $\lambda \in [0, \lambda^* (X;Y))$ with $\delta (t, 0) =0$.
Let
\[ \Delta^* (t) \defeq \lim_{\lambda \uparrow \lambda^*(X;Y)} \delta (t, \lambda) \;.\]
By an argument similar to that in the proof of Theorem~\ref{th1}, it can be shown that $r (t, \delta)$ is strictly increasing, convex and continuously differentiable up to at least the third order over $\delta \in [0, \Delta^* (t))$, and furthermore $r (t, \delta)$ has the following parametric expression
 \begin{equation} \label{eq3rpr}
  r (t, \delta(t, \lambda)) =   \lambda ( I(t; P) + \delta (t, \lambda) ) -
  \sum_{x \in {\cal X}} t(x) \ln  \int  p(y |x) \left [ {p (y|x) \over q_t (y) } \right ]^{\lambda}  d y
   \end{equation}
with
\[ \lambda = {\partial r (t, \delta) \over \partial \delta }\]
 satisfying
 \[ \delta (t, \lambda) = \delta \;.  \]
Further define for any $\lambda \in [0, \lambda^* (X; Y))$
   \begin{equation} \label{eq3rr-13}
  \sigma^2_{D} (t; P, \lambda)  \defeq \sum_{x \in {\cal X}} t (x) \left [ \int p( y|x) f_{\lambda} (y |x)   \left |
  \ln {p(y|x) \over q_t (y) } - D_{+}(t, x, \lambda) \right |^2  d y \right ]
  \end{equation}
  and
\begin{equation} \label{eq3rr-14}
  M_{D} (t; P , \lambda)  \defeq \sum_{x \in {\cal X}} t (x) \left [ \int p( y|x) f_{\lambda} (y |x)   \left |
  \ln {p(y|x) \over q_t (y) } - D_{+}(t, x, \lambda)  \right |^3  d y \right ]
  \;.
  \end{equation}
  Then the following result can be proved similarly, which  is referred to as the right NEP with respect to relative entropy.
\begin{theorem}[Right NEP With Respect to Relative Entropy] \label{th3r+}
For any sequence $x^n=x_1 \cdots x_n$ from $\cal X$, let $t \in {\cal P}$ be the type of $x^n$, i.e., $n t(a)$, $a \in {\cal X}$,  is the number of times the symbol $a$ appears in $x^n$. Assume that $t$ has full support. Then
  \begin{equation} \label{eqrr3-3}
  \Pr \left \{ \left. {1 \over n} \ln {p(Y^n |X^n)  \over q_t (Y^n) } > I(t; P) + \delta \right |  X^n =x^n    \right \} \leq e^{- n r (t, \delta) }\;.
  \end{equation}
 Furthermore,  under the assumptions  \eqref{eqr3-1+} and \eqref{eq3-1+++},  the following also hold:
 \begin{description}

 \item[(a)] There exists a $\delta^* >0$ such that for any $\delta \in (0, \delta^*]$  \begin{equation} \label{eqrr3-4-}
 r  (t, \delta) = {1 \over 2 \sigma^2_D (t; P) } \delta^2 + O(\delta^3)
  \end{equation}
  and hence
 \begin{equation} \label{eqrr3-4}
  \Pr \left \{ \left. {1 \over n} \ln {p(Y^n |X^n)  \over q_t(Y^n) } > I(t; P) + \delta \right | X^n = x^n  \right \} \leq e^{- n ({\delta^2 \over 2\sigma^2_D (t; P) } + O(\delta^3) )}\;.
  \end{equation}

  \item[(b)] For any $\delta \in (0, \Delta^* (X; Y))$
 \begin{eqnarray} \label{eqrr3-16}
  \underline{\xi}_{D}(t;P,\lambda,n)  e^{- n r (t, \delta)} &\leq&
  \Pr \left \{ \left. {1 \over n} \ln {p(Y^n |X^n)  \over q_t(Y^n) } > I(t; P) + \delta \right | X^n = x^n  \right \} \nonumber\\
  &\leq& \bar{\xi}_{D}(t;P,\lambda,n)  e^{- n r(t, \delta)}
  \end{eqnarray}
  where $\lambda = {\partial  r (t, \delta)  \over \partial \delta} >0$, and
  \begin{eqnarray}
    \label{eqrr3-17-1}
    \lefteqn{\bar{\xi}_{D}(t;P,\lambda,n) = \frac{2CM_{D}(t;P,\lambda)}{\sqrt{n} \sigma^3_{D}(t;P,\lambda)}  } \nonumber \\
                      &&{+}\: e^{\frac{n \lambda^2 \sigma^2_{D}(t;P,\lambda)}{2}} 
                      \left[ Q(\sqrt{n}  \lambda \sigma_{D}(t;P,\lambda)) - Q(\rho^*+\sqrt{n}  \lambda \sigma_{D}(t;P,\lambda))\right] \\
    \label{eqrr3-17-2}
   \lefteqn{ \underline{\xi}_{D}(t;P,\lambda,n) = 
       e^{\frac{n \lambda^2 \sigma^2_{D}(t;P,\lambda)}{2}} Q(\rho_*+\sqrt{n}  \lambda \sigma_{D}(t;P,\lambda))}
  \end{eqnarray}
  with $Q(\rho^*) = \frac{CM_D(t;P,\lambda)}{\sqrt{n} \sigma^3_D(t;P,\lambda)}$ and 
  $Q(\rho_*) = \frac{1}{2} - \frac{2CM_D(t;P,\lambda)}{\sqrt{n} \sigma^3_D(t;P,\lambda)}$.

  \item[(c)] For any  $ \delta \leq c \sqrt{\ln n \over n} $, where $c < \sigma_D (t; P)$ is a constant,
  \begin{eqnarray} \label{eqrr3-17+}
    Q  \left ( {\delta \sqrt{n} \over \sigma_D (t; P)} \right ) - {C M_D (t; P) \over \sqrt{n} \sigma^3_D (t; P)}
    & \leq  &   \Pr \left \{ \left. {1 \over n} \ln {p(Y^n |X^n)  \over q_t(Y^n) } > I(t; P) + \delta \right | X^n = x^n  \right \}
    \nonumber \\
  &  \leq &  Q  \left ( {\delta \sqrt{n} \over \sigma_D (t; P)} \right ) + {C M_D (t; P) \over \sqrt{n} \sigma^3_D (t; P)}\;.
     \end{eqnarray}

  \end{description}
\end{theorem}
Remarks similar to those (Remark \ref{remark1} and \ref{remark2}) following Theorem \ref{th1+}
can be drawn here concerning Theorems \ref{th3r} and \ref{th3r+}.
  Theorem~\ref{th3r} will be used in \cite{jar_decoding} to establish a non-asymptotic coding theorem for Shannon random codes.

\section{NEP Application to Fixed Rate Source Coding}
\label{sec4}
\setcounter{equation}{0}

Assume that  the source alphabet $\mathcal{X}$ is finite. In this section, we make use of the NEP with respect to $H(X)$ to establish a non-asymptotic fixed rate source coding theorem, which reveals, for any finite block length $n$, a complete picture about the tradeoff between the minimum rate of fixed rate coding of $X_1 \cdots X_n$ and error probability when the error probability is a constant, or goes to $0$ with block length $n$ at a sub-polynomial $n^{-\alpha}$, $0 < \alpha <1$, polynomial $n^{-\alpha}$, $ \alpha \geq 1$, or sub-exponential $e^{- n^{\alpha}}$, $0< \alpha <1$, speed. We begin with the definition of fixed rate source code.
\begin{definition}
  Given a source from alphabet $\mathcal{X}$, a fixed rate source code
  with coding length $n$ is defined as a mapping $i: S_n \rightarrow
  \{1,2,\ldots,|S_n|\}$, where $S_n$ is a subset of
  $\mathcal{X}^n$. The performance of the code is measured by the rate
  $R_n= \frac{1}{n} \ln |S_n|$ (in nats) and error probability $\Pr
  \left\{ X^n \notin S_n \right\}$.
\end{definition}
As can be seen from the definition, the design of a fixed rate source
code is equivalent to picking a subset of $\mathcal{X}^n$. Given the
source statistics $p(x)$, one can
easily show that the optimal way to pick $S_n$ is to order $x^n$ in the non-increasing order of $p(x^n)$, 
and include those $x^n$ with rank less than or
equal to $|S_n|$. Then we have the following non-asymptotic fixed rate source coding theorem. 

\begin{theorem}
  \label{thm-sc-redun}
  
  Let $R_n (\epsilon_n)$ denote the minimum rate (in nats) of fixed rate coding of $X_1 X_2 \cdots X_n$ subject to the error probability not larger than $\epsilon_n$. Under the assumptions \eqref{eq1+} and \eqref{eq1+++}, for any $n$ and $\epsilon_n >0$,
  \begin{equation}
    \label{eq-thm-sc-1}
    \bar{\delta} \geq  R_n (\epsilon_n) - H(X) \geq \underline{\delta} - r_X (\underline{\delta}) 
     + \frac{- d + \ln \left[ \frac{1}{2} - Q \left( \frac{d}{\sqrt{n} \sigma_H (X,\lambda)}
     \right) -  \frac{2 C M_H (X,
      \lambda)}{\sqrt{n} \sigma^3_H (X, \lambda)}  \right]}{n} 
  \end{equation}
  for any constant $d$ satisfying $\frac{1}{2} - Q \left( \frac{d}{\sqrt{n} \sigma_H (X,\lambda)}
     \right) -  \frac{2 C M_H (X,
      \lambda)}{\sqrt{n} \sigma^3_H (X, \lambda)} >0$,
  where $\bar{\delta}$ is the solution to the equation
  \begin{equation}
    \label{eq-thm-sc-2}
    \epsilon_n = \bar{\xi}_H (X,r'_X (\delta),n) e^{-nr_X(\delta)}
  \end{equation}
  $\underline{\delta}$ is the solution to the equation
  \begin{equation}
    \label{eq-thm-sc-2+}
    \left( 1+ e^{-n} \right) \epsilon_n =  \underline{\xi}_H (X,r'_X (\delta),n)  e^{-n r_X (\delta)}
  \end{equation}
  and $\lambda = r'_X (\underline{\delta}) $.  In particular, the following hold, depending on whether $\epsilon_n$ is a constant, or how fast $\epsilon_n$ goes to $0$. 
  \begin{enumerate}
  \item[(a)] When $\epsilon_n$ decreases exponentially with respect to $n$,
    \begin{eqnarray}
      \label{eq-thm-sc-5}
      r^{(inv)}_X \left( - \frac{\ln \epsilon_n}{n}
      - \frac{\ln
      n}{2n} \right)  + O(n^{-1})
      &\geq& R_n (\epsilon_n) - H(X) \nonumber \\
      &\geq& r^{(inv)}_X \left( - \frac{\ln \epsilon_n}{n}
      - \frac{\ln
      n}{2n} \right)+ \frac{\ln \epsilon_n}{n} - O(n^{-1}) \nonumber \\
    \end{eqnarray}
    where $r^{(inv)}_X (\cdot) $ is the inverse function of $r_X (\cdot) $.
  \item[(b)] When $\epsilon_n = n^{-\frac{\alpha}{2}} e^{-n^{\alpha}}$ for $\alpha \in (0,1)$,
    \begin{eqnarray}
      \label{eq-thm-sc-5+}
      \sqrt{2} \sigma_H (X) n^{-\frac{1-\alpha}{2}} + O \left( n^{-\frac{1+\alpha}{2}} \right) 
      &\geq& R_n(\epsilon_n) - H(X) \nonumber \\
      &\geq& \sqrt{2} \sigma_H (X) n^{-\frac{1-\alpha}{2}} - O \left( n^{-\frac{1+\alpha}{2}} \right)
    \end{eqnarray}
     for $\alpha \in \left( 0, \frac{1}{3} \right)$, and
     \begin{eqnarray}
      \label{eq-thm-sc-5++}
      \sqrt{2} \sigma_H (X) n^{-\frac{1-\alpha}{2}} + O \left( n^{-(1-\alpha)} \right) 
      &\geq& R_n(\epsilon_n) - H(X) \nonumber \\
      &\geq& \sqrt{2} \sigma_H (X) n^{-\frac{1-\alpha}{2}} - O \left( n^{-(1-\alpha)} \right)
    \end{eqnarray}
    for $\alpha \in \left[ \frac{1}{3} , 1 \right)$.
  \item[(c)] When $\epsilon_n = \frac{n^{-\alpha}}{\sqrt{\ln n}}$ for $\alpha > 0$, 
  \begin{eqnarray}
    \label{eq-thm-sc-3}
    \sigma_H (X) \sqrt{\frac{2 \alpha \ln n}{n}} + O \left( \sqrt{\frac{1}{n \ln n}} \right)
        &\geq& R_n (\epsilon_n) - H(X) \nonumber \\
        &\geq& \sigma_H (X) \sqrt{\frac{2 \alpha \ln n}{n}} - O \left( \sqrt{\frac{1}{n \ln n}} \right).
  \end{eqnarray}
   
  \item[(d)]  When $\epsilon_n = \epsilon$ remains a constant,
  \begin{eqnarray}
    \label{eq-thm-sc-7}
    \frac{\sigma_H}{\sqrt{n}} Q^{-1} \left( \epsilon - \frac{C M_H(X)}{\sqrt{n} \sigma^3_H(X)} \right)
    &=& \frac{\sigma_H}{\sqrt{n}} Q^{-1} \left( \epsilon \right) + O \left( \frac{1}{n} \right) \nonumber \\
    &\geq&  R_n (\epsilon_n) - H(X)  \nonumber \\
    &\geq& \frac{\sigma_H}{\sqrt{n}} Q^{-1} \left( \epsilon \right) - O \left( \frac{\ln n}{n} \right).
  \end{eqnarray}
  where $Q^{-1} \left( \cdot \right)$ is the inverse function of $Q \left( \cdot \right)$. 
  \end{enumerate}
\end{theorem}

\begin{IEEEproof}[Proof of Theorem~\ref{thm-sc-redun}]
  Define
  \begin{displaymath}
    S_n (\delta) \defeq \left\{ x^n: - \frac{1}{n} \ln p(x^n) \leq H(X) + \delta \right\}
  \end{displaymath}
  and
  \begin{displaymath}
    \epsilon_n (\delta) = \Pr \left\{ X^n \notin S_n (\delta) \right\} .
  \end{displaymath}
  Clearly $\epsilon_n (\delta)$ is a non-increasing function of $\delta$.
  Now let $\bar{\delta}$ and $\underline{\delta}$ satisfy that
  \begin{equation}
    \label{eq-thm-sc-proof--2}
    \epsilon_n (\bar{\delta}) \leq \epsilon_n < \epsilon_n ( \underline{\delta} ) .
  \end{equation}
  According to the discussion on optimal fixed-rate source codes,
  \begin{equation}
    \label{eq-thm-sc-proof--1}
    \frac{1}{n} \ln S_n (\underline{\delta}) < R_n (\epsilon_n) \leq \frac{1}{n} \ln S_n (\bar{\delta}).
  \end{equation}
  Observe that
  \begin{eqnarray*}
    |S_n (\bar{\delta}) | e^{-n(H(X) + \bar{\delta})} &\leq& \sum_{x^n \in S_n (\bar{\delta}) } p(x^n) \\
    &\leq& \sum_{x^n \in \mathcal{X}^n} p(x^n) \\
    &\leq& 1
  \end{eqnarray*}
  which implies that
  \begin{equation}
    \label{eq-thm-sc-proof-0}
    R_n(\epsilon_n) \leq \frac{1}{n} \ln |S_n (\bar{\delta}) | \leq H(X) + \bar{\delta} .
  \end{equation}
  Towards the lower bound on $R_n(\epsilon_n)$, further define
  \begin{displaymath}
    S_n (\underline{\delta},d) \defeq \left\{ x^n: H(X) + \underline{\delta} - \frac{d}{n}
      \leq - \frac{1}{n} \ln p(x^n) \leq H(X) + \underline{\delta} \right\}
  \end{displaymath}
  for some constant $d > 0$. Then we have
  \begin{eqnarray*}
    |S_{n} (\underline{\delta},d)| e^{- n \left( H(X) + \underline{\delta} - \frac{d}{n}
      \right)} &\geq& \sum_{x^n \in S_n (\underline{\delta}, d)} p(x^n) \\
    &=&  \sum_{x^n \in S_n (\underline{\delta}, d)} f^{-1}_{\lambda} (x)
    f_{\lambda} (x^n) p(x^n) \\
    &=& \sum_{x^n \in S_n (\underline{\delta}, d)} e^{-n \left[ -\frac{1}{n}
        \lambda \ln p(x^n) - \ln \sum_{u \in \mathcal{X}}
        p^{-\lambda+1}(u) \right]} f_{\lambda} (x^n) p(x^n) \\
    &\geq& e^{-n r_X (\underline{\delta})} \sum_{x^n \in S_{n} (\underline{\delta}, d)}
    f_{\lambda} (x^n) p(x^n) \\
    &=& e^{-n r_X (\underline{\delta})} \Pr \left\{ Z^n \in S_n (\underline{\delta},d)
    \right\} \\
    &=& e^{-n r_X (\underline{\delta})} 
    \Pr \left\{ - \frac{d}{n}
      \leq \frac{1}{n} \sum^n_{i=1} -\ln p(Z_i) -(H(X) + \underline{\delta}) \leq 0
    \right\} \\
    &\geq& e^{-n r_X(\underline{\delta})} \left[
     \frac{1}{2} - Q \left( \frac{d}{\sqrt{n} \sigma_H (X,\lambda)}
     \right) -  \frac{2 C M_H (X,
      \lambda)}{\sqrt{n} \sigma^3_H (X, \lambda)} \right]
  \end{eqnarray*}
  where $\lambda = r'_X (\underline{\delta})$, $\{Z_i \}_{i=1}^n$ are IID random variables with common pmf $f_{\lambda} (z) p(z)$,   and the last inequality is due to the direct application of Lemma \ref{le1} (Berry-Esseen Central Limit Theorem)  to $\{ - \ln p(Z_i) -(H(X) + \underline{\delta} ) \}_{i=1}^n$.
  And therefore
  \begin{eqnarray}
    \label{eq-thm-sc-proof-1}
    R_n (\epsilon_n) &>& \frac{1}{n} \ln |S_n(\underline{\delta})| \nonumber \\
    &\geq& \frac{1}{n} \ln |S_n(\underline{\delta}, d)| \nonumber \\
    &\geq& H(X) + \underline{\delta} - \frac{d}{n} - r_X
    (\underline{\delta}) \nonumber \\
    &&{+}\: \frac{1}{n} \ln \left[ \frac{1}{2} - Q \left( \frac{d}{\sqrt{n} \sigma_H (X,\lambda)}
     \right) -  \frac{2 C M_H (X,
      \lambda)}{\sqrt{n} \sigma^3_H (X, \lambda)}  \right] .
  \end{eqnarray}
  Note that $\frac{1}{2} - Q \left( \frac{d}{\sqrt{n} \sigma_H (X,\lambda)}
     \right) -  \frac{2 C M_H (X,
      \lambda)}{\sqrt{n} \sigma^3_H (X, \lambda)} = \Theta \left(
      \frac{1}{\sqrt{n}}\right)$ for constant $d>0$.
  Then \eqref{eq-thm-sc-1} is proved by showing $\bar{\delta}$ and $\underline{\delta}$ calculated according
  to \eqref{eq-thm-sc-2} and \eqref{eq-thm-sc-2+} indeed satisfy \eqref{eq-thm-sc-proof--2}, where we invoke
  Theorem \ref{th1+}, i.e.
  \begin{eqnarray*}
    \epsilon_n (\bar{\delta}) &=& \Pr \left\{ X^n \notin S_n (\bar{\delta}) \right\} \\
                                        &\leq& \bar{\xi}_H (X,r'_X (\bar{\delta}),n) e^{- n r_X (\bar{\delta})} \\
                                        &=& \epsilon_n 
  \end{eqnarray*}
  while
  \begin{eqnarray*}
    \epsilon_n (\underline{\delta}) &=& \Pr \left\{ X^n \notin S_n (\underline{\delta}) \right\} \\
                                        &\geq& \underline{\xi}_H (X,r'_X (\underline{\delta}),n) e^{- n r_X (\underline{\delta})} \\
                                        &>& \epsilon_n.
  \end{eqnarray*}
  
  Let us now look at special cases. 
  \begin{enumerate}
  \item[(a)] When $\epsilon_n$ decreases exponentially with respect to $n$, i.e.
  $\frac{1}{n} \ln \epsilon_n \to c$ as $n \to +\infty$ for some constant $c < 0$,
  we have
    \begin{equation}
      \label{eq-thm-sc-proof-a1}
     \frac{\ln \epsilon_n}{n} = \frac{\ln \bar{\xi}_H (X,r'_X (\bar{\delta}),n)}{n} - r_X (\bar{\delta}).
    \end{equation}
  Note that
  \begin{displaymath}
    \bar{\xi}_H(X,\lambda,n) \geq \frac{2 C M_H(X,\lambda)}{\sqrt{n} \sigma^3_H(X,\lambda)} = \Omega \left( \frac{1}{\sqrt{n}} \right).
  \end{displaymath}
  Taking $n \to +\infty$ in \eqref{eq-thm-sc-proof-a1}, it can be seen that $r_X(\bar{\delta}) \to -c$. And therefore,
  $\bar{\xi}_H (X,r'_X (\bar{\delta}),n) = \Theta \left( \frac{1}{\sqrt{n}} \right)$, which further implies that
  \begin{eqnarray}
    \label{eq-thm-sc-proof-a1+}
    \bar{\delta} &=& r^{(inv)}_X \left( -\frac{\ln \epsilon_n}{n} + \frac{\ln \bar{\xi}_H (X,r'_X (\bar{\delta}),n)}{n}  \right) \nonumber \\
                      &= & r^{(inv)}_X \left( -\frac{\ln \epsilon_n}{n} - \frac{\ln n}{2n}  + O(n^{-1}) \right) \nonumber \\
                      &=& r^{(inv)}_X \left( -\frac{\ln \epsilon_n}{n} - \frac{\ln n}{2n}  \right) + O(n^{-1}).
  \end{eqnarray}
  On the other hand,
  \begin{equation}
    \label{eq-thm-sc-proof-a2}
        \frac{\ln \epsilon_n}{n} + \frac{\ln (1+e^{-n})}{n} = 
              \frac{\ln \underline{\xi}_H (X,r'_X (\underline{\delta}),n)}{n} - r_X (\underline{\delta}).
  \end{equation}
  and by the same argument, $r_X(\underline{\delta}) \to -c$ as $n \to +\infty$. Consequently,
  $\underline{\xi}_H (X,r'_X (\underline{\delta}),n) = \Theta \left( \frac{1}{\sqrt{n}} \right)$, which further implies
  \begin{equation}
    \label{eq-thm-sc-proof-a3}
    \frac{\ln \epsilon_n}{n} = -r_X (\underline{\delta}) - \frac{\ln n}{2n} + O(n^{-1})
  \end{equation}
  and
  \begin{equation}
    \label{eq-thm-sc-proof-a4}
    \underline{\delta} = r^{(inv)}_X \left( -\frac{\ln \epsilon_n}{n} - \frac{\ln n}{2n}  \right) - O(n^{-1}).
  \end{equation}
  Combining \eqref{eq-thm-sc-1} with \eqref{eq-thm-sc-proof-a1+}, 
  \eqref{eq-thm-sc-proof-a3} and \eqref{eq-thm-sc-proof-a4} yields,
  \begin{eqnarray}
    \label{eq-thm-sc-proof-a5}
    r^{(inv)}_X \left( -\frac{\ln \epsilon_n}{n} - \frac{\ln n}{2n}  \right) + O(n^{-1})
    &\geq& R_n(\epsilon_n) - H(X) \nonumber \\
    &\geq& \underline{\delta} - r_X(\underline{\delta}) - \frac{\ln n}{2n} - O(n^{-1}) \nonumber \\
    &= & r^{(inv)}_X \left( -\frac{\ln \epsilon_n}{n} - \frac{\ln n}{2n}  \right) + \frac{\ln \epsilon_n}{n} - O(n^{-1}) \nonumber \\
  \end{eqnarray}
 This completes the proof of \eqref{eq-thm-sc-5}.
  \item[(b)] First of all, let us consider the case when $\alpha \in \left( 0, \frac{1}{3} \right)$. 
    Towards proving \eqref{eq-thm-sc-5+}, 
    let us show that $\bar{\delta} = \sqrt{2} \sigma_H (X) n^{-\frac{1 - \alpha}{2}} + \eta n^{-\frac{1+\alpha}{2}}$ for some
    properly chosen constant $\eta$ will guarantee
    \begin{equation}
      \label{eq-thm-sc-proof-a+1}
      \epsilon_n (\bar{\delta}) \leq n^{-\frac{\alpha}{2}} e^{-n^{\alpha}}. 
    \end{equation}
    By Theorem \ref{th1+} and Remark \ref{remark1},
    \begin{displaymath}
      \epsilon_n \left( \bar{\delta} \right)
      \leq \bar{\xi}_H \left( X,r'_X \left( \bar{\delta} \right), n \right)
             e^{- n r_X \left( \bar{\delta} \right)}
    \end{displaymath}
    while
    \begin{eqnarray*}
      \lefteqn{\bar{\xi}_H \left( X,r'_X \left( \bar{\delta} \right), n \right)} \\
        &=& \bar{\xi}_H \left( X,
                                        r'_X \left( \sqrt{2} \sigma_H (X) n^{-\frac{1 - \alpha}{2}} + \eta n^{-\frac{1+\alpha}{2}} \right), 
                                        n \right) \\
        &=& \Theta \left( 
                                   \frac{1}
                                          {\sqrt{n} r'_X \left( \sqrt{2} \sigma_H (X) n^{-\frac{1 - \alpha}{2}} 
                                           + \eta n^{-\frac{1+\alpha}{2}} \right) 
                                          } 
                           \right) \\
      &=& \Theta \left( n^{-\frac{\alpha}{2}} \right) \leq \eta_1 n^{-\frac{\alpha}{2}}
    \end{eqnarray*}
     for some constant $\eta_1 > 0$, and
     \begin{eqnarray*}
       \lefteqn{e^{-n r_X (\bar{\delta})}} \\
         &=& \exp \left\{ - n r_X \left( \sqrt{2} \sigma_H (X) n^{-\frac{1 - \alpha}{2}} + \eta n^{-\frac{1+\alpha}{2}} \right) \right\} \\
         &=& \exp \left\{ - n \left[ \frac{1}{2 \sigma^2_H(X) }
                                                \left( \sqrt{2} \sigma_H (X) n^{-\frac{1 - \alpha}{2}} + \eta n^{-\frac{1+\alpha}{2}} \right)^2  
                                                +  O \left( n^{-\frac{3(1-\alpha)}{2}} \right) 
                                        \right] \right\} \\
         &=& \exp \left\{ - n^{\alpha} - \frac{\sqrt{2} \eta}{\sigma_H(X)} - O \left( n^{-\alpha} + n^{-\frac{1-3 \alpha}{2} }\right) \right\} \\
         &=& \exp \left\{ - n^{\alpha} - \frac{\sqrt{2} \eta}{\sigma_H(X)} - o(1) \right\}
     \end{eqnarray*}
    since $\alpha \in \left(0, \frac{1}{3} \right)$. Now it is trivial to see that we can select a constant $\eta$ such that
    \begin{displaymath}
      \eta_1 e^{- \frac{\sqrt{2} \eta}{\sigma_H(X)} - o(1)} \leq 1
    \end{displaymath}
    which will make \eqref{eq-thm-sc-proof-a+1} satisfied, and consequently
    \begin{eqnarray*}
      \bar{\delta}
      &=& \sqrt{2} \sigma_H (X) n^{-\frac{1 - \alpha}{2}} + \eta n^{-\frac{1+\alpha}{2}} \\
      &=& \sqrt{2} \sigma_H (X) n^{-\frac{1 - \alpha}{2}} +  O \left( n^{-\frac{1+\alpha}{2}} \right) \\
      &\geq& R_{n} (\epsilon_n) - H(X).
    \end{eqnarray*}
    In the similar manner, we can show that
    by making $\underline{\delta} = \sqrt{2} \sigma_H (X) n^{-\frac{1 - \alpha}{2}} - \eta' n^{-\frac{1+\alpha}{2}}$ for
    another constant $\eta' > 0$,
    \begin{displaymath}
      \epsilon_n (\underline{\delta}) > \epsilon_n.
    \end{displaymath}
    Consequently,
    \begin{eqnarray*}
      R_n (\epsilon_n) - H(X) 
       &\geq& \underline{\delta} - r_X (\underline{\delta}) - \frac{\ln n}{2 n} - O(n^{-1}) \\
       &=& \sqrt{2} \sigma_H (X) n^{-\frac{1 - \alpha}{2}} - \eta' n^{-\frac{1+\alpha}{2}} - O \left( n^{-(1-\alpha)} \right) \\
       &=& \sqrt{2} \sigma_H (X) n^{-\frac{1 - \alpha}{2}} - O \left( n^{-\frac{1+\alpha}{2}} \right)
    \end{eqnarray*}
    for $\alpha \in \left( 0, \frac{1}{3} \right)$. 
    The proof of \eqref{eq-thm-sc-5++} for the case $\alpha \in \left[ \frac{1}{3}, 1 \right)$ is essentially the same, 
    and therefore omitted.
  \item[(c)] Following the same spirit of the proof for part (b), one can verify that
  constants $\eta$ and $\eta'$ can be chosen respectively such that
  \begin{displaymath}
    \left[ \epsilon_n \left( \bar{\delta} \right) \left|_{\bar{\delta} =  \sigma_H \sqrt{\frac{2 \alpha \ln n}{n}} + \eta \sqrt{\frac{1}{n \ln n}} } \right. \right]
    \leq \frac{n^{-\alpha}}{\sqrt{\ln n}}
  \end{displaymath}
  and
  \begin{displaymath}
    \left[ \epsilon_n \left( \underline{\delta} \right) 
            \left|_{\underline{\delta} =  \sigma_H \sqrt{\frac{2 \alpha \ln n}{n}} - \eta' \sqrt{\frac{1}{n \ln n}} } \right.
    \right]
    > \frac{n^{-\alpha}}{\sqrt{\ln n}}
  \end{displaymath}
  which, together with \eqref{eq-thm-sc-1}, proves \eqref{eq-thm-sc-3}.
  \item[(d)]
   It can be readily seen that by Theorem \ref{th1+} (b),
   $\bar{\delta} = \frac{\sigma_H (X)}{\sqrt{n}} Q^{-1} \left( \epsilon - \frac{C M_H(X)}{\sqrt{n} \sigma^3_H(X)} \right)$
   is the right choice to guarantee
   \begin{displaymath}
     \epsilon_n (\bar{\delta}) \leq \epsilon
   \end{displaymath}
   while $\underline{\delta} = \frac{\sigma_H (X}{\sqrt{n}} Q^{-1} \left( \epsilon + \frac{2 C M_H(X)}{\sqrt{n} \sigma^3_H(X)} \right)$
   will make
   \begin{displaymath}
     \epsilon_n (\underline{\delta}) > \epsilon
   \end{displaymath}
   satisfied. \eqref{eq-thm-sc-7} then follows immediately from \eqref{eq-thm-sc-1} and  the choices of $\bar{\delta}$ and $\underline{\delta}$.
   \end{enumerate}
  This completes the proof of Theorem~\ref{thm-sc-redun}.
  \end{IEEEproof}

\begin{remark} \em
  To show Theorem \ref{thm-sc-redun} provides a non-trivial bound, we
  claim that
  \begin{displaymath}
    \delta > r_X (\delta)
  \end{displaymath}
  for $0 < \delta < \ln |\mathcal{X}| - H(X)$.
  Indeed, recall the definition of $\delta (\lambda)$ and
  \begin{displaymath}
    0 \leq r_X(\delta(1)) = H(X) + \delta(1) - \ln |\mathcal{X}|
  \end{displaymath}
  which implies that $\delta(1) \geq \ln |\mathcal{X}| - H(X)$ or
  $r'_X (\delta) < 1$ for $0 < \delta < \ln |\mathcal{X}| -
  H(X)$. The claim then follows immediately from the fact that $r_X (0) =0$. 
\end{remark}

\begin{figure}[h]
  \centering
  \includegraphics[scale=0.5]{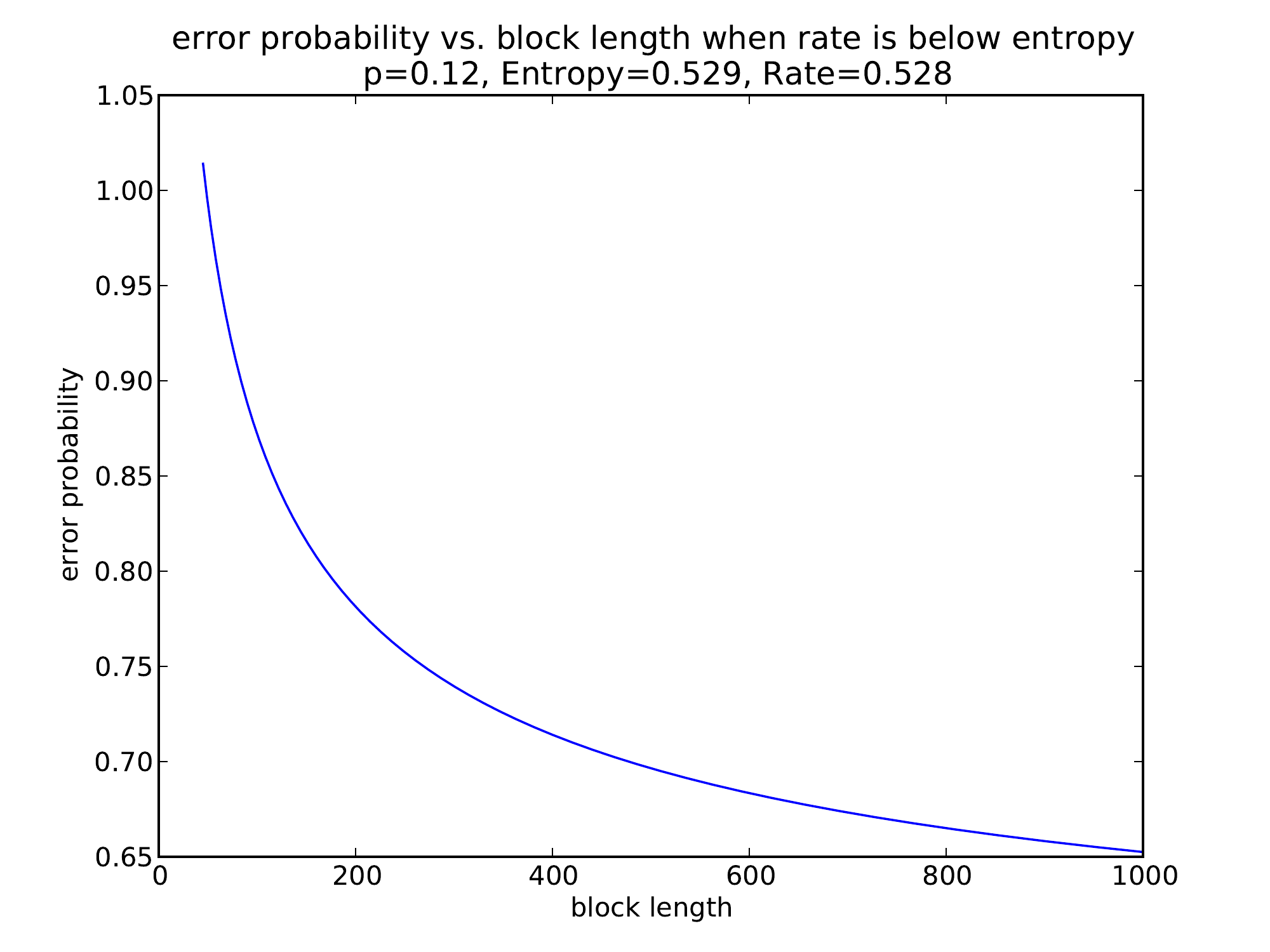}
  \caption{Tradeoff between the error probability and block length
    when the rate is below the entropy rate with $p=0.12$}
  \label{fig-blelow-entrate}
\end{figure}

\begin{remark} \em
  In Part (d) of Theorem \ref{thm-sc-redun}, we can see that if $\epsilon_n =\epsilon > 0.5$ is selected,
  then $R_n (\epsilon_n) $ could be strictly less than $H(X)$  for finite block length $n$! This means that
  if the error probability is allowed to be slightly larger than $0.5$,
  the rate of source code can be even less than the entropy rate. For an IID binary source with $p = \Pr \{ X_1 =1 \} = 0.12$,  Figure~\ref{fig-blelow-entrate}  
  shows the tradeoff between the error
      probability and block length when the code rate is $0.21\%$
      below the entropy rate, where in Figure~\ref{fig-blelow-entrate}, both the entropy rate and code rate are
      expressed in terms of bits. As can be seen from
      Figure~\ref{fig-blelow-entrate},  at the block length $1000$, the 
      error probability is around $0.65$, and the code rate is
      $0.21\%$ below the entropy rate. 
  Similar
  phenomenon can be seen for channel coding shown in \cite{jar_decoding}.
\end{remark}

\begin{remark} \em Related to Part (d) of Theorem
  \ref{thm-sc-redun} is the 
  second order source coding analysis  in \cite{hayashi} with a fixed error
  probability $0 < \epsilon < 1$. Both results are concerned
  with the scenario where the rate is around the entropy rate in the order
  of $\frac{1}{\sqrt{n}}$ and the error probability is a
  constant. However, the work in \cite{hayashi} is asymptotic. On the other hand, 
   Theorem \ref{thm-sc-redun} (\eqref{eq-thm-sc-1} and Part (d)) is non-asymptotic and valid for any block length $n$. It reveals a complete picture about
  the tradeoff between the rate and error probability when the error
  probability is constant, or approaches $0$ with block length $n$ at an exponential
  (Part (a)), a sub-exponential (Part (b)),  a polynomial (Part (c) with $\alpha \geq 1$), or a sub-polynomial (Part (c) with $0< \alpha <1$) speed.
\end{remark}

\begin{thebibliography}{99}


\bibitem{Cover}
T. M. Cover and J. A. Thomas, {\em Elements of Information Theory (second edition)}.
Hoboken, NJ: Wiley, 2006.

\bibitem{hall} P. Hall, {\em Rates of Convergence in the Central Limit
Theorem.} Pitman Books Limited, Boston, 1982.

\bibitem{jar_decoding}
E. -H. Yang and J. Meng, ``Jar decoding: Basic Concepts and Non-asymptotic Capacity Achieving Coding Theorems for Channels with Discrete Inputs,'' {\em in preparation for submission to IEEE Trans. Inform. Theory} for publications, 2011.

\bibitem{yang-redundancy} E.-H. Yang and Zhen Zhang, ``On the
  redundancy of lossy source coding with abstract alphabets,''  {\em
    IEEE Trans. Inform. Theory}, vol. 44, pp. 1092--1110, May 1999.

\bibitem{hayashi} M. Hayashi, ``Second order asymptotics in fixed-length source
  coding and intrinsic randomness'', {\em IEEE Trans. Inform. Theory},
  vol. 54, pp. 4619 - 4637, Oct. 2008. 
\end{thebibliography}
\end{document}